\documentclass[preprint,5p,times,twocolumn]{elsarticle}
\usepackage{amsmath}
\usepackage{hyperref}
\usepackage{graphicx,amssymb,lineno}
\usepackage{epstopdf}
\usepackage{amsthm}
\usepackage{mathrsfs}
\usepackage{float}
\usepackage{subfigure}
\usepackage{array}
\usepackage{booktabs}
\usepackage{longtable}
\usepackage{txfonts}
\usepackage{caption}
\usepackage{multirow}
\usepackage{algorithm}
\usepackage{algorithmic}
\usepackage{threeparttable}
\usepackage{eqnarray}
\usepackage{color,xcolor}
\usepackage{bm}

\newcommand{\pref}[1]{{\rm(\ref{#1})}}
\newcommand\norm[1]{\left\Vert#1 \right\Vert}

\journal{}

\begin{document}
\baselineskip11pt

\begin{frontmatter}

\title{Reasonable thickness determination for implicit porous sheet structure using persistent homology}

\author[label1,label2]{Jiacong Yan}
\author[label1,label2]{Hongwei Lin\corref{cor1}}
\ead{hwlin@zju.edu.cn}
\cortext[cor1]{Corresponding author.}

\address[label1]{School of Mathematical Science, Zhejiang University, Hangzhou, 310027, China}
\address[label2]{State Key Laboratory of CAD\&CG, Zhejiang University, Hangzhou, 310058, China}

\begin{abstract}
  Porous structures are widely used in various industries because of their excellent properties.
  Porous surfaces have no thickness and should be thickened to sheet structures for further fabrication.
  However,
    conventional methods for generating sheet structures are inefficient for porous surfaces because of the complexity of the internal structures.
  In this study,
    we propose a novel method for generating porous sheet structures directly from point clouds sampled on a porous surface.
  The generated sheet structure is represented by
    an implicit B-spline function,
    which ensures smoothness and closure.
  Moreover, based on the persistent homology theory,
    the topology structure of the generated porous sheet structure can be controlled,
    and a reasonable range of the uniform thickness of the sheet structure can be calculated to ensure manufacturability and pore existence.
  Finally,
    the implicitly B-spline represented sheet structures are sliced directly with the marching squares algorithm,
    and the contours can be used for 3D printing.
  Experimental results show the superiority of the developed method in efficiency over the traditional methods.
\end{abstract}


\end{frontmatter}

\section{Introduction}
\label{sec:intro}

 Porous structures are solid structures with many pores.
 Compared with continuous medium structures,
    porous structures have many attractive properties,
    such as low relative density, high specific mechanical strength,
    and large internal surface area \cite{fang2005computer}.
 Therefore, porous structures are widely used in mechanical, tissue,
    and chemical engineering \cite{ajdari2011dynamic,yoo2011porous,chen2018metal}.
 Porous structures can be manufactured effectively
    owing to developments in 3D printing technologies
    \cite{hu2021heterogeneous,zhai2019path,feng2019layered}.
 However, porous surfaces without thickness cannot be
    manufactured directly and should be thickened into
    solid structures in advance.
 The mechanical properties of porous sheet structures,
    generated by inflating porous surfaces to a predefined finite
    constant thickness,
    have significant advantages over conventional designs \cite{kapfer2011minimal}.
 Hence, the methods for generating porous sheet structures for 3D printing are
    worthy of investigation.

 In general, the types of representations for porous surface are
    explicit representation,
    such as mesh representation,
    and implicit representation.
 On the one hand, for mesh represented surface \cite{yoo2009general,liu2010fast,wang2013thickening},
    the sheet structure is generated by initially
    calculating two offset surfaces,
    and then sealing them to form a closed model for manufacturing.
 However, the offset operation is complicated
    and error prone \cite{park2005hollowing},
    especially to porous surfaces with complicated geometric shape
    and topology structure.
 On the other hand, with an implicitly B-spline represented porous surface,
    such as $g(x,y,z)=a$,
    although generating a sheet structure is easier by
    extracting the region
    $a_1 \leq g(x,y,z) \leq a_2$, where $a_1 \leq a_2$ \cite{hu2020efficient},
    the thickness of the sheet is varied
    and difficult to control directly.
 This condition will affect the mechanical properties of
    the generated sheet structure greatly.

 In this study, we develop a method for generating
    implicitly B-spline represented porous sheet structures
    with a uniform thickness from point clouds.
 Specifically, given a point cloud sampled from a porous surface,
    we initially calculate a discrete unsigned distance field (DUDF) to the point cloud model,
    and fit the DUDF with a trivariate B-spline function, i.e., $f(x,y,z)$.
 Then, the porous sheet structure is implicitly B-spline represented as
    $f(x,y,z) \leq c$,
    where $c$ is the \emph{thickness parameter},
    with $2c$ as its \emph{thickness}.
 First, because the trivariate B-spline function is a distance field,
    the thickness of the generated sheet structure is uniform.
 Secondly, because the distance field
    is unsigned,
    the generated sheet structure is naturally closed.
 Finally, the implicitly B-spline represented sheet structure saves great storage space,
    and is easy to slice directly using marching squares (MS) algorithm \cite{maple2003geometric} for 3D printing.

 However, the thickness parameter $c$ of the
    implicitly B-spline represented porous
    sheet structure should be determined deliberately.
 On the one hand, if the thickness parameter $c$ is extremely small,
    the extra small holes can possibly emerge on the surface of
    the porous sheet structure.
 On the other hand, if the thickness of the sheet structure
    is extremely large,
    the pores will be closed.
 We develop a method for determining the thickness by
    understanding the topology structure of the distance field represented by
    a trivariate B-spline function using \emph{persistent homology} \cite{edelsbrunner2000topological}.
 Persistent homology is a powerful tool for recording topological features
    and evaluating their importance using \emph{persistent diagram} (PD) in a \emph{filtration} procedure.
 In our implementation, we produce the PDs with sub-level filtration of the trivariate B-spline function,
    and the PDs reveal the topology structure of the distance field.
 By clustering the points in the PDs,
    we can obtain a reasonable range of thickness,
    which can not only avoid the pores from closing,
    but also eliminate small extra holes in the porous sheet structure.
 To our best knowledge,
    this algorithm is the first to calculate the reasonable thickness range,
    which maintains all the pores open,
    in the sheet structure generation.

 In summary, the main contributions of this work are as follows:
 \begin{itemize}
    \item We develop a method for generating an implicitly B-spline represented porous sheet structure from point cloud by fitting a DUDF,
        which saves great storage space and is easy to slice for fabrication.
    \item The implicitly B-spline represented porous sheet structure has a uniform thickness and is naturally closed given the properties of DUDF.
    \item A reasonable thickness range of the sheet structure is determined using persistent homology,
        which can maintain all the pores in the porous sheet structure open
        and eliminate extra small holes.
 \end{itemize}

 The remainder of the paper is organized as follows.
 In Section \ref{sec:related},
    we review related works on implicit surface reconstruction, sheet structure generation and slicing methods, and the application of computational topology.
 In Section \ref{sec:preliminaries},
    preliminaries on the cubical complex and persistent homology are introduced.
 In Section \ref{sec:sheet},
    we propose an effective method for generating porous sheet structures with implicit B-spline representations and determine their reasonable thickness range.
 In Section \ref{sec:experimental},
    some experimental results are presented to illustrate the effectiveness of the proposed algorithm.
 Finally, the paper is concluded in Section \ref{sec:conclusion}.

\section{Related work}
\label{sec:related}

 In this section, we list related works on implicit surface reconstruction,
    sheet structure generation and slicing methods,
    and the application of computational topology.

\subsection{Implicit surface reconstruction from point clouds}
\label{subsec:surface}

 The basic idea of implicit surface reconstruction from point clouds is representing the surface as a zero-level set of a scalar function or scalar field.
 Hoppe et al. \cite{hoppe1992surface} proposed the signed distance field reconstruction method,
    which is one of the earliest implicit surface reconstruction methods from point clouds.
 Furthermore,
    different types of functions are used to approximate distance functions for extracting zero-level sets.
 Carr et al. \cite{carr2001reconstruction} presented using radial basis functions to model a large data set.
 Compactly supported radial basis functions were used in \cite{morse2005interpolating} for reducing computation effort.
 Spline functions are also commonly used in surface reconstruction.
 Wang et al. \cite{wang2011parallel} generated implicit PHT-spline surfaces from the given point clouds,
    which are over hierarchical T-meshes.
 Hamza et al. \cite{hamza2020implicit} proposed an implicit surface reconstruction algorithm called I-PIA based on B-spline basis functions.
 The I-PIA algorithm eliminates the extra zero level set and is robust to low quality point clouds.

 In addition to the distance function, the indicator function has been used for surface reconstruction.
 Kazhdan \cite{kazhdan2005reconstruction} used Fourier series to represent indicator function for reconstructing surface from points equipped with oriented normals.
 To improve the Fourier series approach,
    the Poisson reconstruction method \cite{kazhdan2006poisson} reconstructed triangle meshes from point clouds by transforming the surface reconstruction problem into a Poisson problem to approximate the indicator function and extract the iso-surface.
 Kazhdan and Hoppe \cite{kazhdan2013screened} developed the typical Poisson reconstruction method to prevent over-smoothing by incorporating the positional constraints in the optimization problem,
    and proposed the screened Poisson surface reconstruction method.
 However, the open surfaces without thickness generated by these algorithms are indirectly manufacturable.

\subsection{Sheet structure generation and slicing}
\label{subsec:sheet}

 The mesh representation, such as triangular mesh, is adopted as the de facto industry standard input \cite{marsan1997survey,pandey2003slicing}.
 Yoo \cite{yoo2009general} introduced a method to generate sheet models based on the nonuniform offset of triangular meshes.
 Liu and Wang \cite{liu2010fast} proposed a method to generate offset surfaces based on the closed triangular mesh model and preserve the sharp features of the model.
 Wang and Chen \cite{wang2013thickening} presented a thickening operation to transform intersection-free mesh surfaces into sheet-like structures and presented the results with signed distance functions.
 However, sealing open triangular meshes is difficult and prone to self-intersection and invalid triangles.
 Many slicing methods are based on triangular meshes \cite{McMains1999285,huang2002effective,zhao2009computing}.
 A large number of facets is required to improve the accuracy of the conversion of an initial model to a triangular mesh.
 The possible errors in the triangular mesh include voids, overlaps, and self-intersections.
 These conditions lead to calculation problems in slicing.
 Therefore, sheet structure and slicing methods based on the triangular mesh are unsuitable for porous structures.

 Implicit representation more easily represents models with complex structures and implements Boolean operations than the mesh representation;
    this condition is interesting to 3D printing.
 Many direct slicing algorithms for implicit representation \cite{song2018function,hong2021direct,popov2020efficient} based on the MS algorithm have been proposed.
 These slicing algorithms are faster and more accurate than slicing algorithms based on mesh representation.
 When an implicit function is expressed for a surface,
    solid structures can be generated by combining two disjoint iso-surfaces \cite{hu2020efficient}.
 However,
    these methods should modify unclosed boundaries by performing Boolean operations with the closed models and increase the information required to be stored.
 Furthermore,
    the thickness of the generated solid structure cannot be directly controlled and is not uniform,
    resulting in fabricated models without the same geometry as the initial surface.

 Some researchers have proposed various methods for directly generating sheet structures from point clouds to solve problems arising from offsetting based on the above representations.
 Liu and Wang \cite{liu2009duplex} implemented a method to fit zero-level surfaces and internal offset surfaces simultaneously according to point clouds coupled with normals.
 However,
    directly representing the sheet structures of the open surfaces based on the fitting results of this method is potentially erroneous.
 Wang and Manocha \cite{wang2013gpu} generated offset surfaces from point clouds through GPU acceleration.
 The main idea is to compute the union of balls centered on input points.
 However,
    this method relies heavily on the sampling density of point clouds.

\subsection{Application of computational topology}
\label{sec:topology}

 Computational topology has boosted the design techniques based on topological features in recent years.
 For geometric design, topology-controllable modeling has attracted increasing attention.
 This method aims to design geometries that satisfy specific topological features.
 Sharf et al. \cite{sharf2006competing} reconstructed a surface with accurate topology from a point cloud based on genus control.
 Attene et al. \cite{attene2013polygon} proposed a post-processing method for repairing topological errors in mesh represented surfaces.
 Chen et al. \cite{chen2019topological} used persistent homology to design regularization and loss functions that incorporate the importance of topological features.

 In the classification and retrieval of porous structures,
    methods based on computational topology show better results than the traditional geometry-based methods.
 Bubenik et al. \cite{bubenik2015statistical} proposed a topological descriptor for retrieving zeolite structures through sampling points on a porous surface and encoding the topological information into a vector.
 Dong et al. \cite{dong2021multiscale} introduced a multiscale persistent topological descriptor for porous structure retrieval,
    which encodes topological information and geometric information.

 These studies demonstrate the great advantages of computational topology for the analysis of topologically complex porous structures.

\section{Preliminaries}
\label{sec:preliminaries}

 Some key concepts on persistent homology used in the current study are introduced in this section.

 \begin{figure}[!htb]
    \centering	
    \subfigure[]{\includegraphics[width=0.12\textwidth]{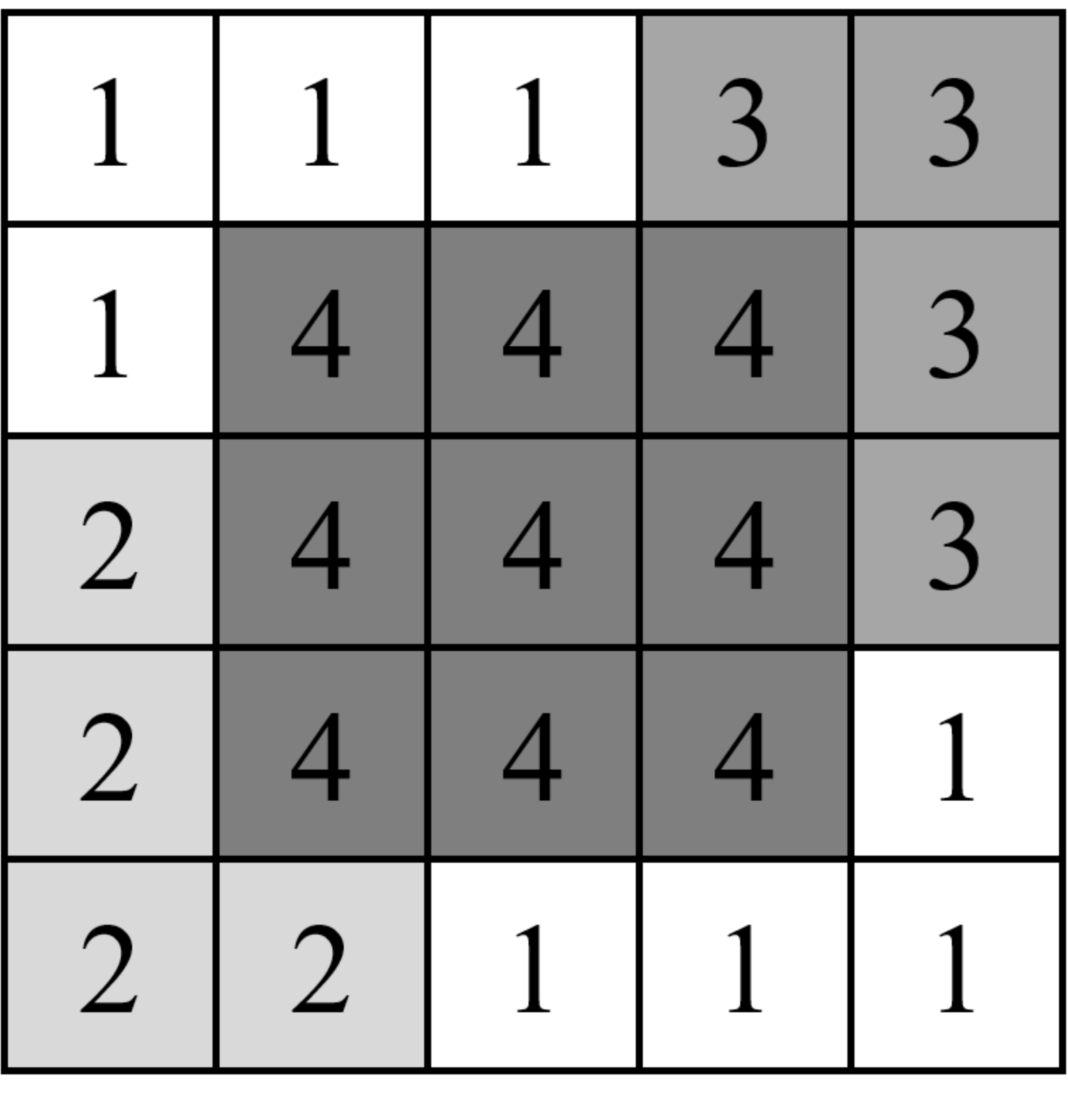}}
    \qquad
    \subfigure[]{\label{subfig:PD} \includegraphics[width=0.15\textwidth]{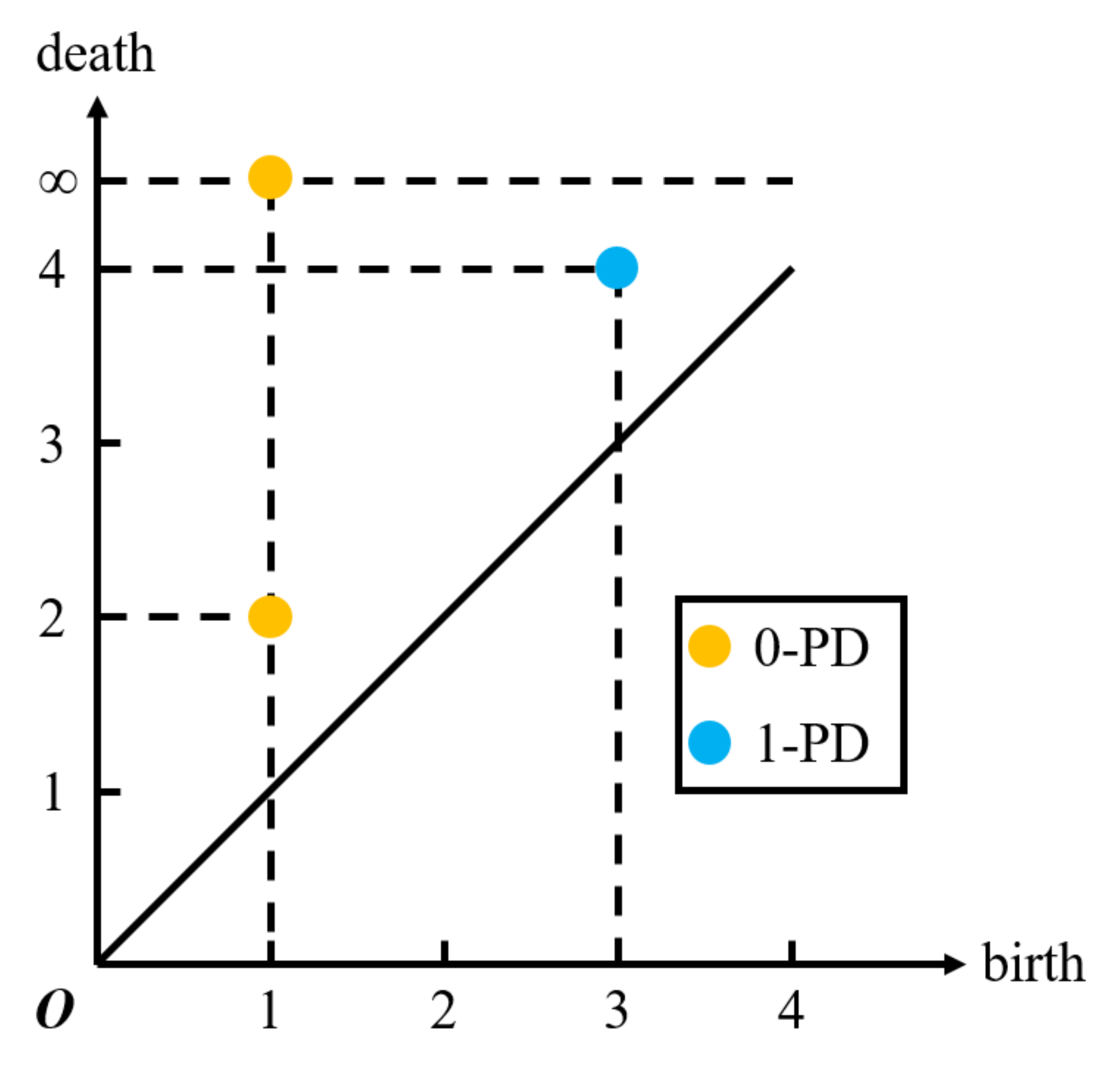}}

    \subfigure[]{\label{subfig:barcode}
    \includegraphics[width=0.36\textwidth]{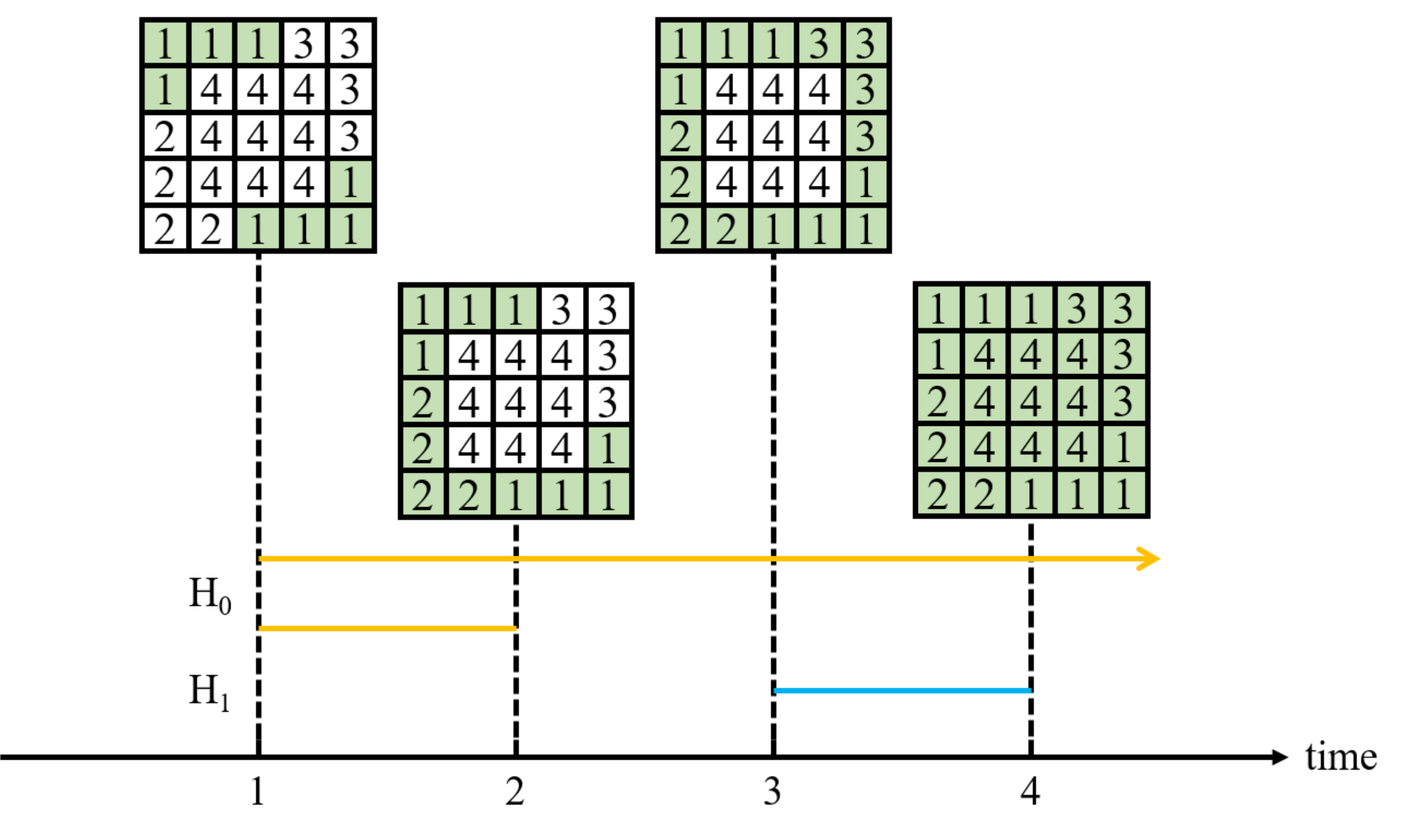}}

    \caption
    {
        Illustration of the sub-level filtration of cubical complex and its corresponding 0, 1-PDs.
        (a) A cubical data with real function values on its grid cells.
        (b, c) An example of the sub-level filtration of cubical complex and PDs.
    }
    \label{fig:PD}
 \end{figure}

\subsection{Cubical complex}
\label{subsec:cubical}

 Cubical complex \cite{kaczynski2004computational} was first used to represent pixel and voxel data,
    which are important and commonly used.
 The elementary cube is the Cartesian product of the elementary interval, such as vertices (0-cubes), line segments (1-cubes), squares (2-cubes), and spatial cubes (3-cubes).
 A cubic complex is a collection of several elementary cubes,
    and the union of all elementary cubes in the cubic complex forms the underlying space of the complex.
 A cubical complex is regarded as the discretization of its underlying space.
 Therefore, a cubic complex is suitable for handling grid-like data,
    such as distance field.

\subsection{Persistent homology}
\label{subsec:persistent}

 Homology is one of the fundamental tools of algebraic topology,
    which can compute and represent the topological features of a given topological space or complex.
 The purpose of homology is to relate a series of Abelian groups and their algebraic relations to the topological space and the relations among the elements in this space.
 By calculating the $k$-th homology group ${H}_{k}(Cub)$ over a given cubic complex $Cub$,
    we can obtain the $k$-dimensional cycles in $Cub$,
    such as connected components, torus, and voids with respect to 0, 1, and 2-dimension.
 $k$-Betti number, defined as the rank of ${H}_{k}(Cub)$,
    represents the number of cycles in the $k$-dimension.

 Furthermore, persistent homology is used for handling a sequence of nested complexes called filtration,
    and capturing the \emph{birth} (or construction) and \emph{death} (or destruction) of topological features in this sequence.
 Sub-level filtration is used in this study.
 Given a real-valued function $f:Cub \rightarrow \mathbb{R}$ defined on the cubic complex $Cub$,
    the cubical subcomplex filtered by $f$ is defined as ${Cub}_{T}=\{\zeta \in Cub | f(\zeta)\leq T\}$,
    where $\zeta$ is an elementary cube,
    and $T\in\mathbb{R}$.
 The sub-level filtration of $Cub$ filtered by $f$ is a nested sequence of cubical subcomplexes that satisfy ${Cub}_{{T}_{i}}\subseteq{Cub}_{{T}_{j}}$ for all ${{T}_{i}}<{{T}_{j}}$ and ${{T}_{i}},{{T}_{j}}\in\mathbb{R}$.
 We can capture the birth times and death times of the cycles of different dimensions in the sub-level filtration as the parameter $T$ increases,
    as illustrated in Fig.~\ref{subfig:barcode}.
 As shown in Fig.~\ref{subfig:PD}, the points $({b}_{i}, {d}_{i})$ embedded into ${\mathbb{R}}^{2}$,
    called \emph{persistence pairs},
    are used as the representation of the topological features in $Cub$,
    where ${b}_{i}$ and ${d}_{i}$ are the birth and death times of the corresponding topological features respectively.
 This most common visualization in persistent homology is called persistent diagram.
 The collection of persistence pairs representing the $k$-dimensional cycles is called $k$-PD.
 ${d}_{i}-{b}_{i}$ is the \emph{persistence time} of the topological feature.

\section{Sheet structure generation and thickness range determination}
\label{sec:sheet}

\begin{figure*}[!htb]
    \centering	
    \subfigure[]{\includegraphics[width=0.175\textwidth]{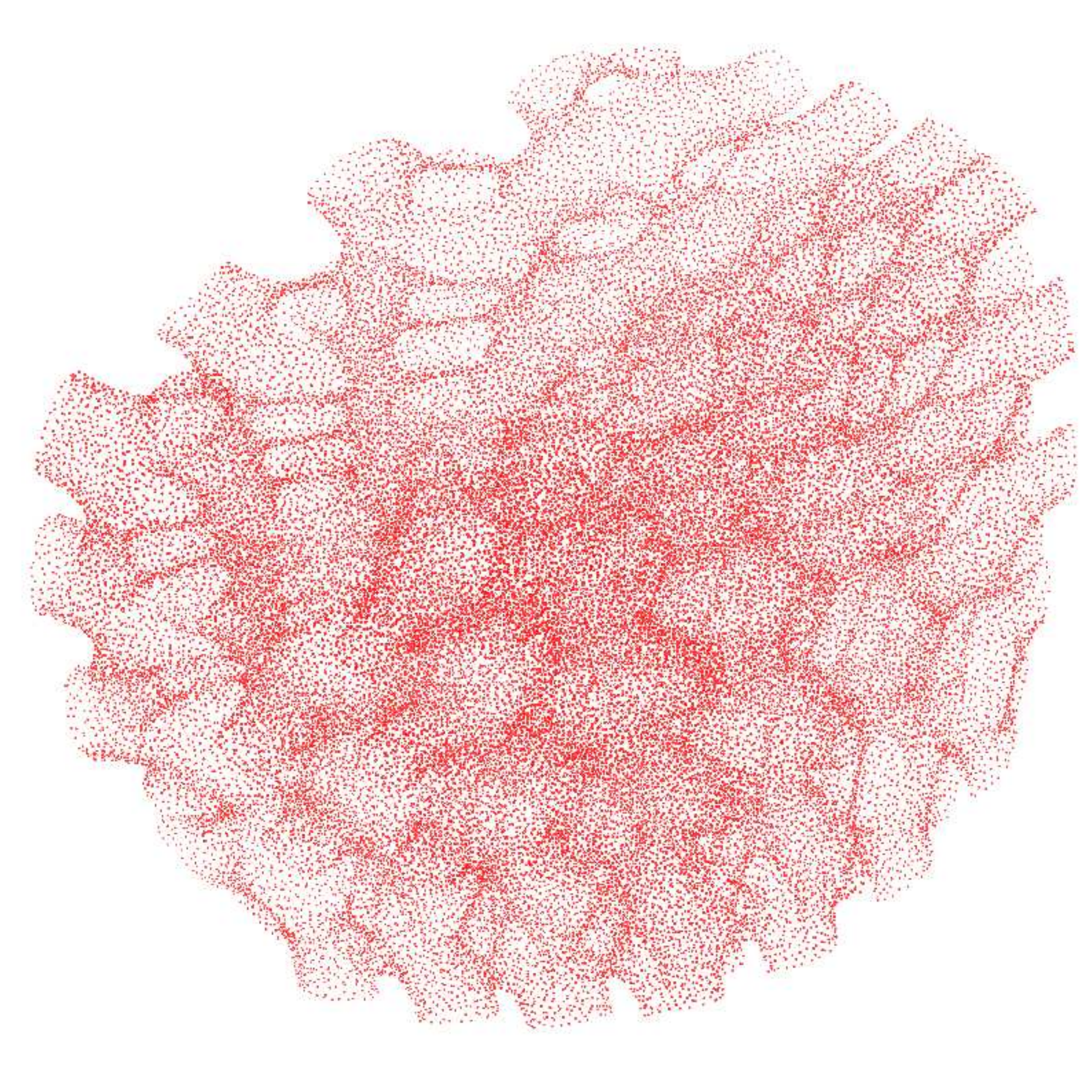}}
    \quad
    \subfigure[]
    {\label{subfig:distance}
        \includegraphics[width=0.175\textwidth]{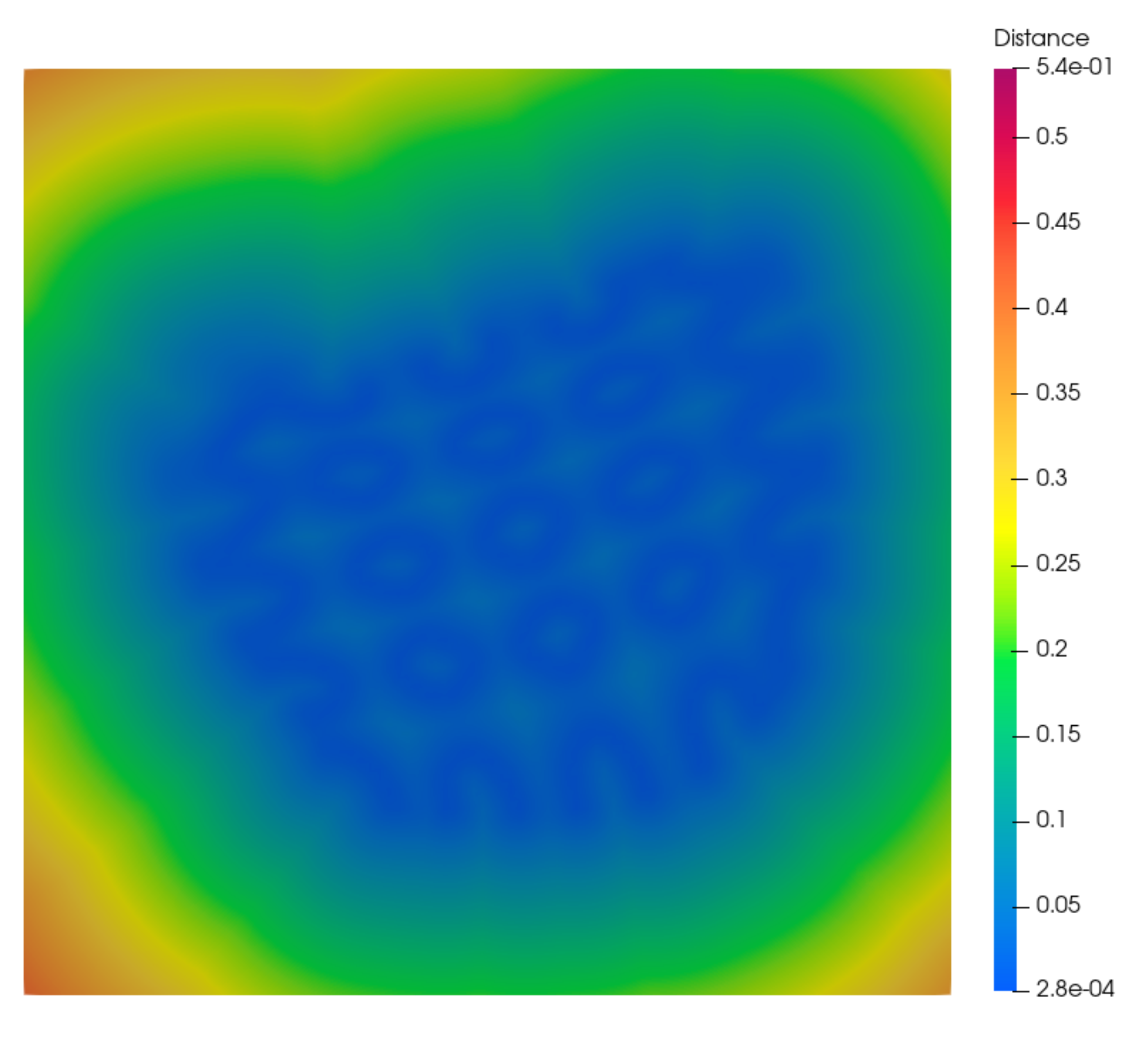}
    }
    \quad
    \subfigure[]
    {\label{subfig:tooth p scalar}
        \includegraphics[width=0.175\textwidth]{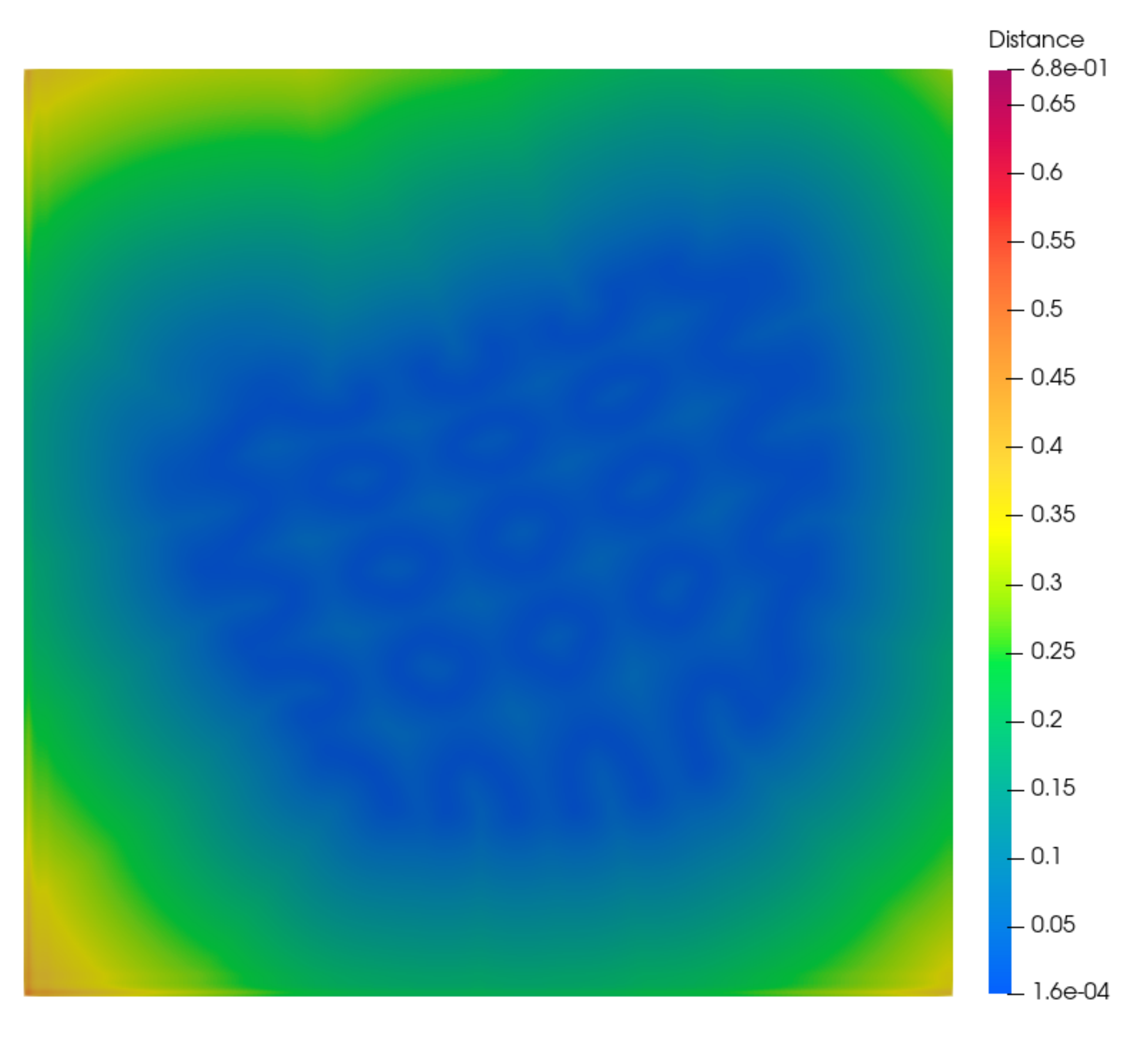}
    }
    \quad
    \subfigure[]{\includegraphics[width=0.175\textwidth]{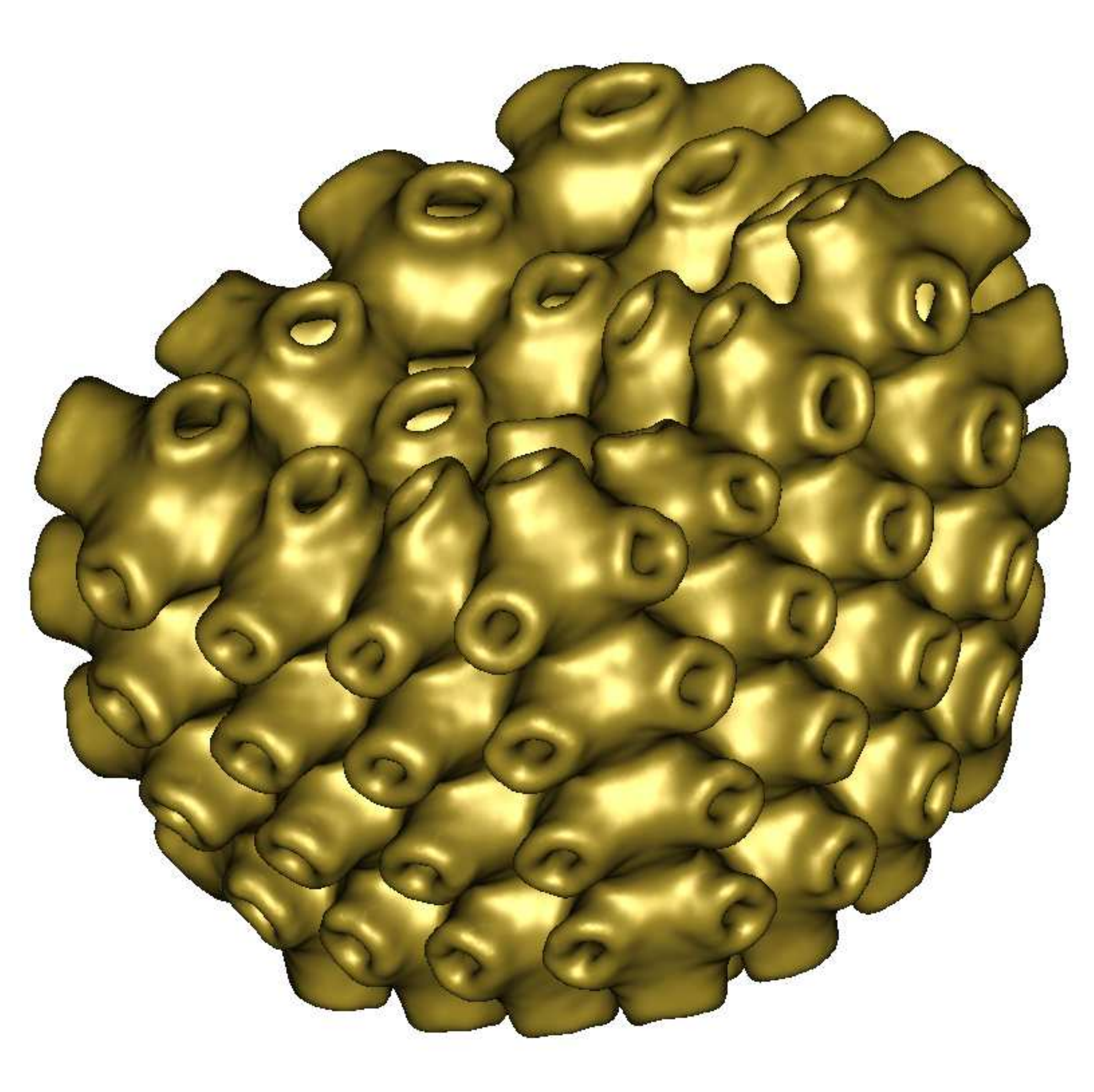}}
    \quad
    \subfigure[]{\includegraphics[width=0.175\textwidth]{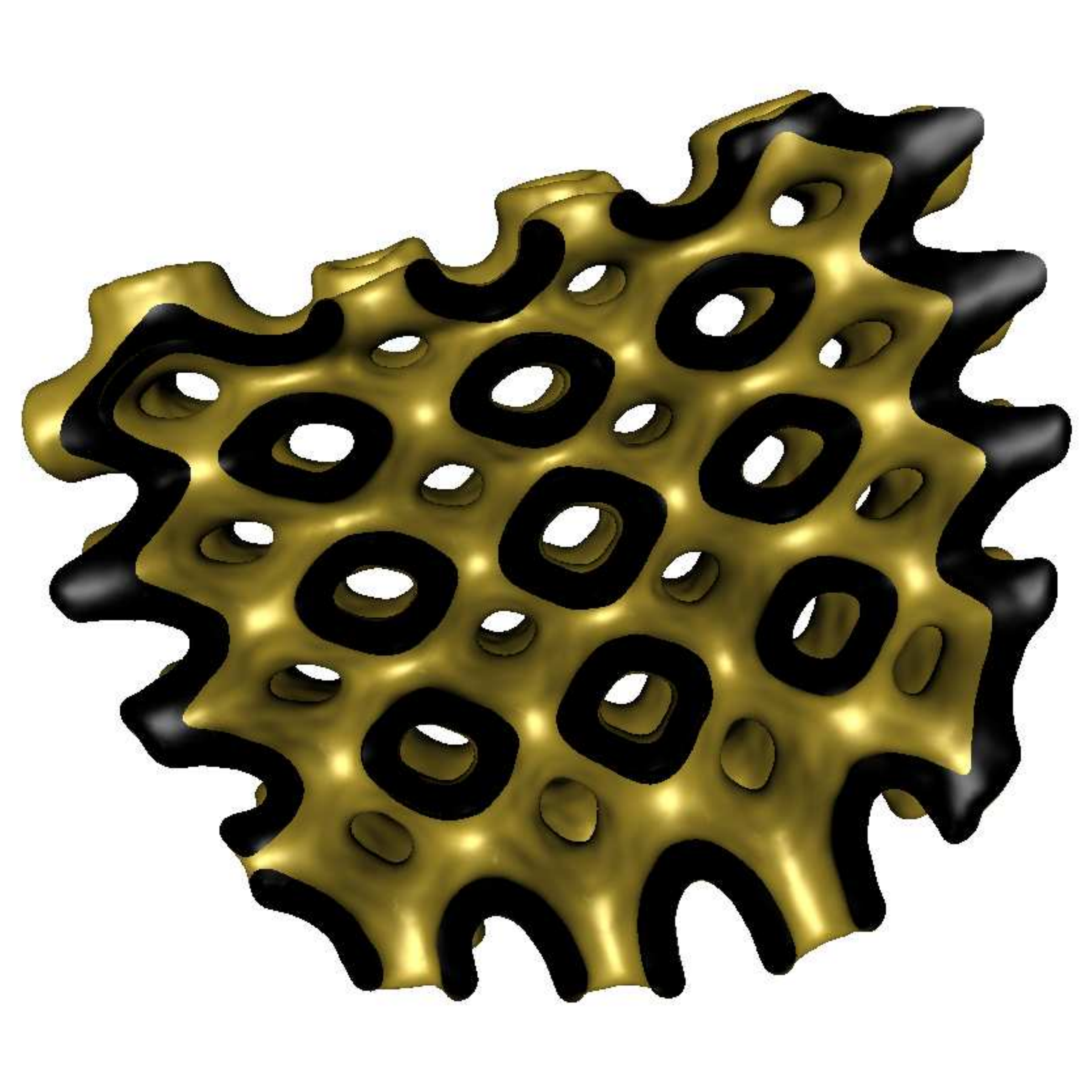}}

    \caption
    {
        Illustration of the proposed porous sheet structure generation method.
        (a) Point cloud (\emph{Tooth P}) with 71,933 points.
        (b) The cutaway view of DUDF with the resolution of $100 \times 100 \times 100$ generated from the input point cloud.
        (c) The cutaway view of scalar field with the resolution of $200 \times200 \times 200$ of trivariate B-spline function $f(x,y,z)$ with a
            $128\times128\times128$ control grid generated by fitting the distance field.
        (d, e) Implicitly B-spline represented \emph{Tooth P} sheet structure $f(x,y,z) \leq 0.01$ and its cutaway view.
    }
    \label{fig:Illustration}
 \end{figure*}

 In this section,
    we propose a method for generating the porous sheet structure
    with implicit B-spline representation from a given point cloud,
    and determining the reasonable thickness range of the generated sheet structure.
 Initially, given a point cloud sampled from a porous surface,
    we generate a DUDF according to the input point cloud, as shown in Fig.~\ref{subfig:distance}.
 Then,
    the DUDF is fitted using a trivariate B-spline function $f(x,y,z)$.
 Moreover, by calculating the 1-PD of the B-spline function $f(x,y,z)$ and clustering the persistent pairs in the 1-PD,
    we can obtain the topological features representing pores and extra small holes, respectively;
    they will be employed to determine the reasonable thickness range of the generated sheet structure.
 Finally, the porous sheet structure with desired porosity is generated by optimization.
 Fig.~\ref{fig:Illustration} demonstrates the generation process of a porous sheet structure.
 Details of the proposed algorithm are elucidated in the subsequent sections.

\subsection{Sheet structure generation with implicit representation}
\label{subsec:sheet_implicit}

 Given a point cloud sampled from a porous surface $M$, i.e.,
 \begin{equation} \label{eq:data_pnts}
    \mathcal{P}=\{{\boldsymbol{p}_{i}}\in M,i=1,2,\cdots,n\},
 \end{equation}
    we initially determine its axis-aligned bounding box (AABB),
    and the AABB is enlarged $\frac{1}{5}$ along each direction
    to ensure all data points are inside.
 Then, the AABB is discretized into a grid,
 \begin{equation}\label{eq:nodes}
 \begin{aligned}
    \mathcal{G}=\{& {\boldsymbol{g}_{ijk}}=({{x}_{i}},{{y}_{j}},{{z}_{k}}),\
    i=1,2,\cdots, {dist}_{nx},\\
    & j=1,2,\cdots,{dist}_{ny},\
    k=1,2,\cdots,{dist}_{nz}\},
 \end{aligned}
 \end{equation}

 and the DUDF of $\mathcal{P}$ can be defined as
 \begin{equation}
 \begin{aligned}
    {{Dist}_{\mathcal{P}}}= \{ & {{d}_{ijk}}=dist({\boldsymbol{g}_{ijk}},\mathcal{P})=\underset{{\boldsymbol{p}_{l}}\in \mathcal{P}}{\mathop{\inf }}{{\Vert {\boldsymbol{g}_{ijk}}-{\boldsymbol{p}_{l}} \Vert}_{2}}, \\
    & i=1,\cdots,{dist}_{nx}, j=1,\cdots,{dist}_{ny}, k=1,\cdots,{dist}_{nz}\},
 \end{aligned}
 \label{eq:DUDF}
 \end{equation}
 where ${\Vert \cdot \Vert}_{2}$ is the Euclidean norm.

\begin{figure}[!htb]
    \centering
    \subfigure[Signed distance field]{\label{subfig:Offset2} \includegraphics[width=0.2\textwidth]{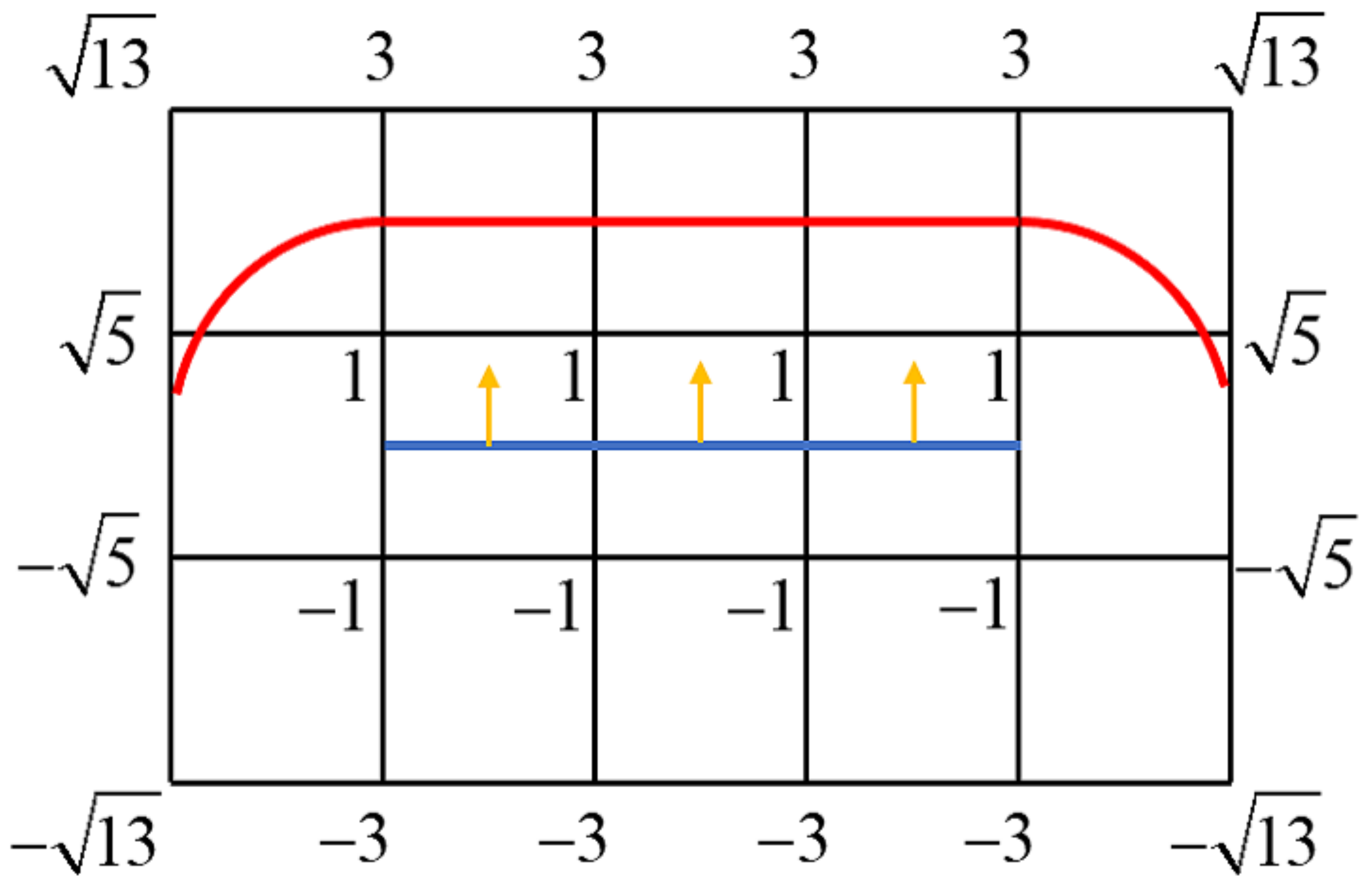}}
    \quad
    \subfigure[Unsigned distance field]{\label{subfig:Offset1} \includegraphics[width=0.2\textwidth]{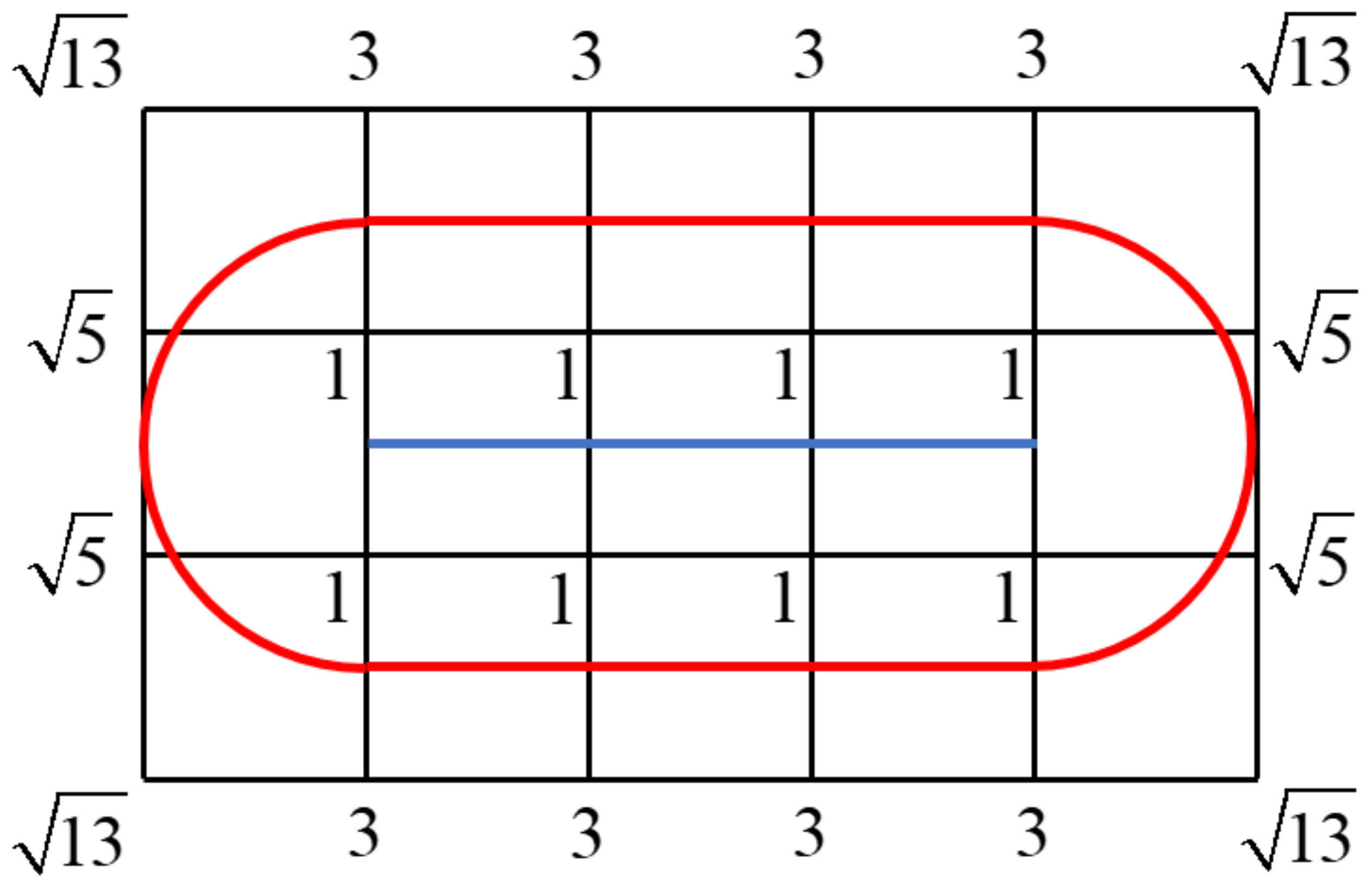}}

    \caption
    {
        Schematic of fitting based on signed distance field (a) and unsigned distance field (b).
        The blue lines and red lines indicate the initial surfaces and the iso-surfaces $f(x,y,z)=c$.
        The yellow arrows indicate normals of the initial surface, respectively.
        When the initial surface is open,
            the iso-surface based on the signed distance field is open in (a),
            and the iso-surface based on the unsigned distance field is closed in (b).
    }
    \label{fig:Offset}
 \end{figure}

 Distance fields include signed and unsigned distance fields.
 In a signed distance field,
    positive and negative distances are calculated according to the unit normal vectors of a model,
    which can be employed in distinguishing between the interior and exterior of the model.
 When fitting is based on signed distance fields,
    the iso-surface $f(x,y,z)=c$ represents the one-sided offset surface because distance values have positive and negative categories.
 Therefore, as shown in Fig.~\ref{subfig:Offset2},
    when the initial surface is an open surface,
    the iso-surface $f(x,y,z)=c$ is open.
 In this case,
    we require extra effort to close boundaries to generate a sheet structure.
 However, as shown in Fig.~\ref{subfig:Offset1},
    when fitting is based on an unsigned distance field,
    the iso-surface $f(x,y,z)=c$ can directly represent the bidirectional offset surface of the initial surface,
    and ensure the closure of the corresponding offset surface.
 Therefore,
    an unsigned distance field is more suitable for generating closed sheet structures.
 Thus, we adopt the unsigned distance field in our work.

 Specifically,
    we calculate the DUDF ${{Dist}_{\mathcal{P}}}$~\pref{eq:DUDF} with the K-d tree~\cite{bentley1975multidimensional}.
 K-d tree~\cite{bentley1975multidimensional} is a binary search tree
    that partitions a data set in the $k$-dimensional space for range search
    and nearest neighbor search.
 Given the data point set $\mathcal{P}$~\pref{eq:data_pnts},
    we can generate the K-d tree based on $\mathcal{P}$ and
    search for the nearest point in $\mathcal{P}$ to a given grid vertex $\bm{g}$.
 The distance at the grid vertex $\bm{g}$ is the distance between $\bm{g}$
    and its nearest point.
%

 After generating the DUDF~\pref{eq:DUDF} of the input point cloud,
    we fit it with a trivariate B-spline function $f(x,y,z)$.
 The trivariate B-spline function $f(x,y,z)$ is defined as
 \begin{equation}
    f(x,y,z)=\sum\limits_{i=0}^{{{N}_{u}}}{\sum\limits_{j=0}^{{{N}_{v}}}{\sum\limits_{k=0}^{{{N}_{w}}}{{{C}_{ijk}}{{B}_{i}}(x){{B}_{j}}(y){{B}_{k}}(z)}}},
    \label{eq:implicit}
 \end{equation}
 where ${{C}_{ijk}}$ are the unknown control coefficients and
 ${{B}_{i}}(x)$, ${{B}_{j}}(x)$, ${{B}_{k}}(x)$
 are B-spline basis functions \cite{Piegl1997the}.
 In this study, we employ the uniform cubic B-spline functions.
 For ease of expression,
    the grid vertices~\pref{eq:nodes} and the DUDF~\pref{eq:DUDF} are rearranged in lexicographic order, i.e.,
    $$ \mathcal{G}=\{{\boldsymbol{g}_{l}}=({{x}_{l}},{{y}_{l}},{{z}_{l}}),\
 l=1,2,\cdots, {dist}_{nx}{dist}_{ny}{dist}_{nz}\},\ \text{and},$$
 $$ {{Dist}_{\mathcal{P}}}= \{ {d}_{l},\ l =1,2,\cdots, {dist}_{nx}{dist}_{ny}{dist}_{nz} \}.$$

 The control coefficients ${{C}_{ijk}}$ can be obtained by solving the least-squares fitting problem
 \begin{equation}
   \min_f \sum_l (d_l - f(\bm{g}_l))^2.
   \label{eq:least-squares}
 \end{equation}

 In this work,
    we adopt the least squares progressive-iteration approximation
    (LSPIA) method~\cite{lin2015constructing,lin2018convergence}
    for solving the least-squares fitting problem~\eqref{eq:least-squares}.
 Suppose that $\alpha $ steps have been performed to obtain the $\alpha $-th trivariate B-spline function ${{f}^{(\alpha )}}(x,y,z)$.
 Let
 \[\delta _{l}^{(\alpha )}={{d}_{l}}-{{f}^{(\alpha )}}({{x}_{l}},{{y}_{l}},{{z}_{l}}),\]
 \[\Delta _{ijk}^{(\alpha )}=\frac{\sum\limits_{l\in {{I}_{ijk}}}{{{B}_{i}}({{x}_{l}}){{B}_{j}}({{y}_{l}}){{B}_{k}}({{z}_{l}})\delta _{l}^{(\alpha )}}}{\sum\limits_{l\in {{I}_{ijk}}}{{{B}_{i}}({{x}_{l}}){{B}_{j}}({{y}_{l}}){{B}_{k}}({{z}_{l}})}}.\]
 the $(\alpha+1)^{th}$ control coefficients and $(\alpha+1)^{th}$
    trivariate B-spline function can be generated as
 \begin{equation}
    C_{ijk}^{(\alpha +1)}=C_{ijk}^{(\alpha )}+\Delta _{ijk}^{(\alpha )},
    \label{eq:coefficients_alpha}
 \end{equation}
 \begin{equation}
    {{f}^{(\alpha +1)}}(x,y,z)=\sum\limits_{i=0}^{{{N}_{u}}}{\sum\limits_{j=0}^{{{N}_{v}}}{\sum\limits_{k=0}^{{{N}_{w}}}{C_{ijk}^{(\alpha +1)}{{B}_{i}}(x){{B}_{j}}(y){{B}_{k}}(z)}}}.
    \label{eq:implicit_alpha}
 \end{equation}

 The above steps are performed iteratively,
    till the stop criteria is satisfied, i.e.,
 \begin{equation*}
    \frac{\norm{\sum_{ijk}(\Delta_{ijk}^{(\alpha+1)}-\Delta_{ijk}^{(\alpha)})}_2}
         {\norm{\sum_{ijk}\Delta_{ijk}^{(0)}}_2} < \varepsilon,
 \end{equation*}
    where $\norm{\cdot}_2$ is the Euclidean norm.

 In our implementation,
    the initial control coefficients $C_{ijk}^{(0)}$ are set to 0,
    and we take $\varepsilon = 10^{-5}$.
 It has been shown that the LSPIA iteration above converges to the solution
    of the least-squares fitting problem
    ~\cite{lin2015constructing,lin2018convergence}.

 \begin{figure*}[!htb]
    \centering
    \subfigure[$c=0.005$]{\label{fig:c range_tooth_P12} \includegraphics[width=0.32\textwidth]{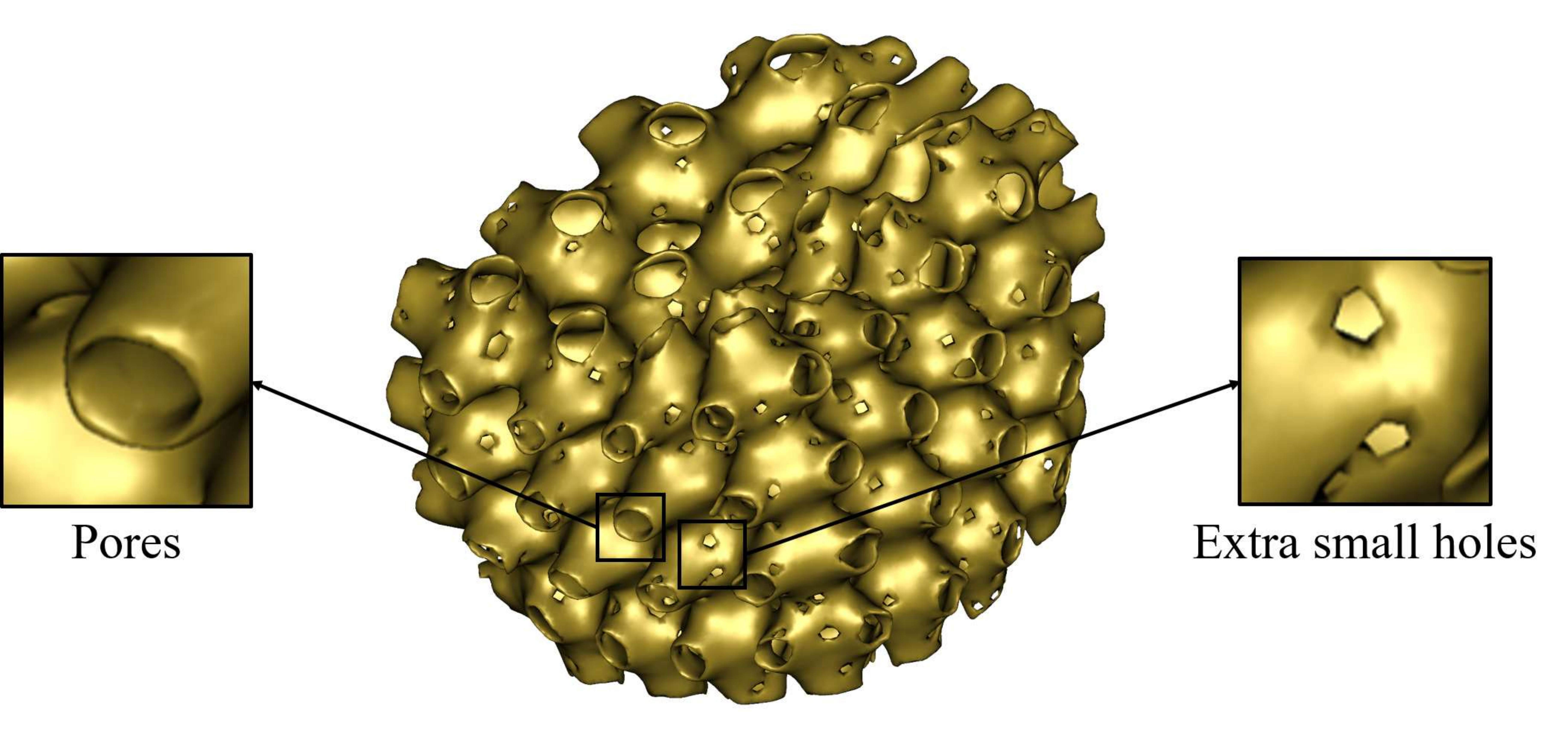}}
    \quad
    \subfigure[$c=0.007$]{\label{fig:c range_tooth_P13} \includegraphics[width=0.15\textwidth]{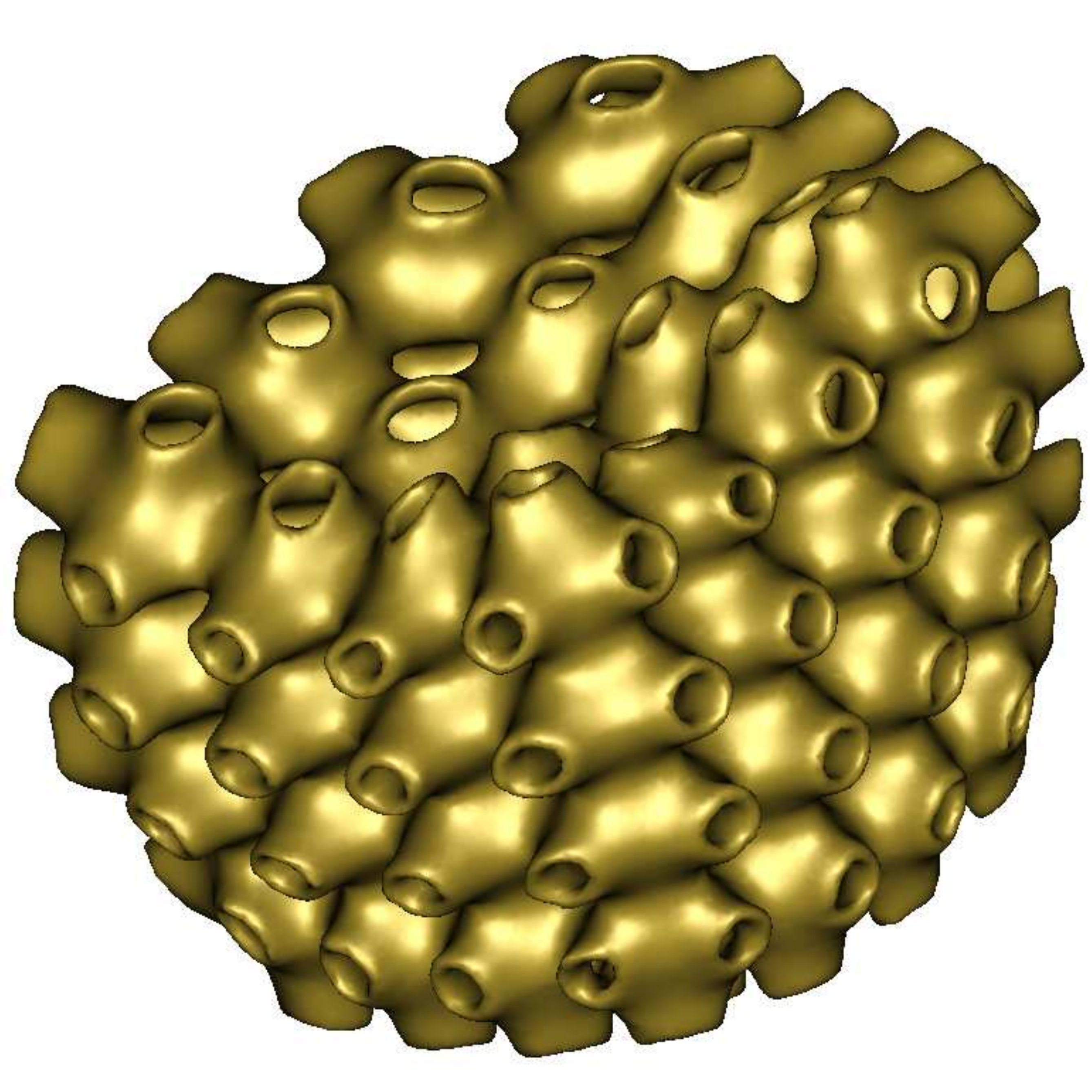}}
    \quad
    \subfigure[$c=0.06$]{\label{fig:c range_tooth_P14} \includegraphics[width=0.15\textwidth]{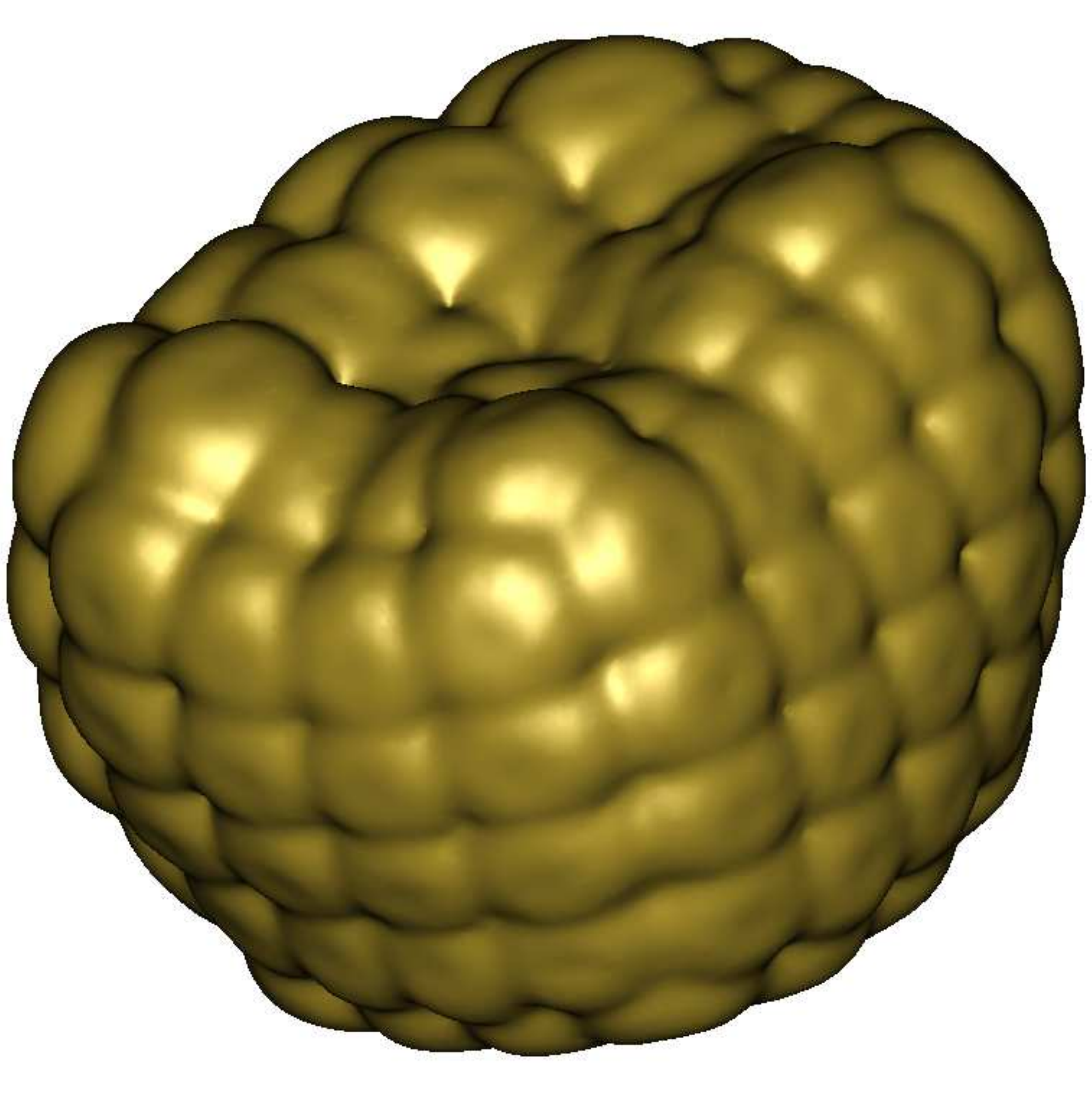}}

    \caption
    {
        \emph{Tooth P} sheet structure $f(x,y,z) \leq c$ extracted from the scalar field \ref{subfig:tooth p scalar} with different thickness parameter $c$.
    }
    \label{fig:c range_tooth_P1}
 \end{figure*}

\subsection{Determination of thickness range}
\label{subsec:offset distance}

 In this study,
    a sheet structure can be represented implicitly as $f(x,y,z) \leq c \ (c>0)$.
 The thickness parameter $c$ represents the bidirectional offset distance of
    an initial surface,
    which is used for controlling the thickness of the sheet structure.
 The thickness parameter $c$ can affect the strength and mass of the
    generated sheet structure monotonically, i.e.,
    strength and mass increase as thickness parameter $c$ increases.

 However, on one hand, when the thickness parameter $c$ is extremely small,
    extra small holes can possibly appear on the sheet structure,
    as illustrated in Fig.~\ref{fig:c range_tooth_P12}.
 On the other hand, as shown in Fig.~\ref{fig:c range_tooth_P14},
    when the thickness parameter $c$ is larger,
    the pores will be closed.
 Therefore, to generate a desirable porous sheet structure,
    a reasonable range of the thickness parameter $c$ should be determined,
    which can maintain all the pores open,
    and eliminate all the extra small holes,
    as illustrated in Fig.~\ref{fig:c range_tooth_P13}.

\begin{figure}[!htb]
    \centering
    \subfigure[1-PD of the distance field $f(x,y,z)$.]{\label{subfig:tooth_P_PD} \includegraphics[width=0.45\textwidth]{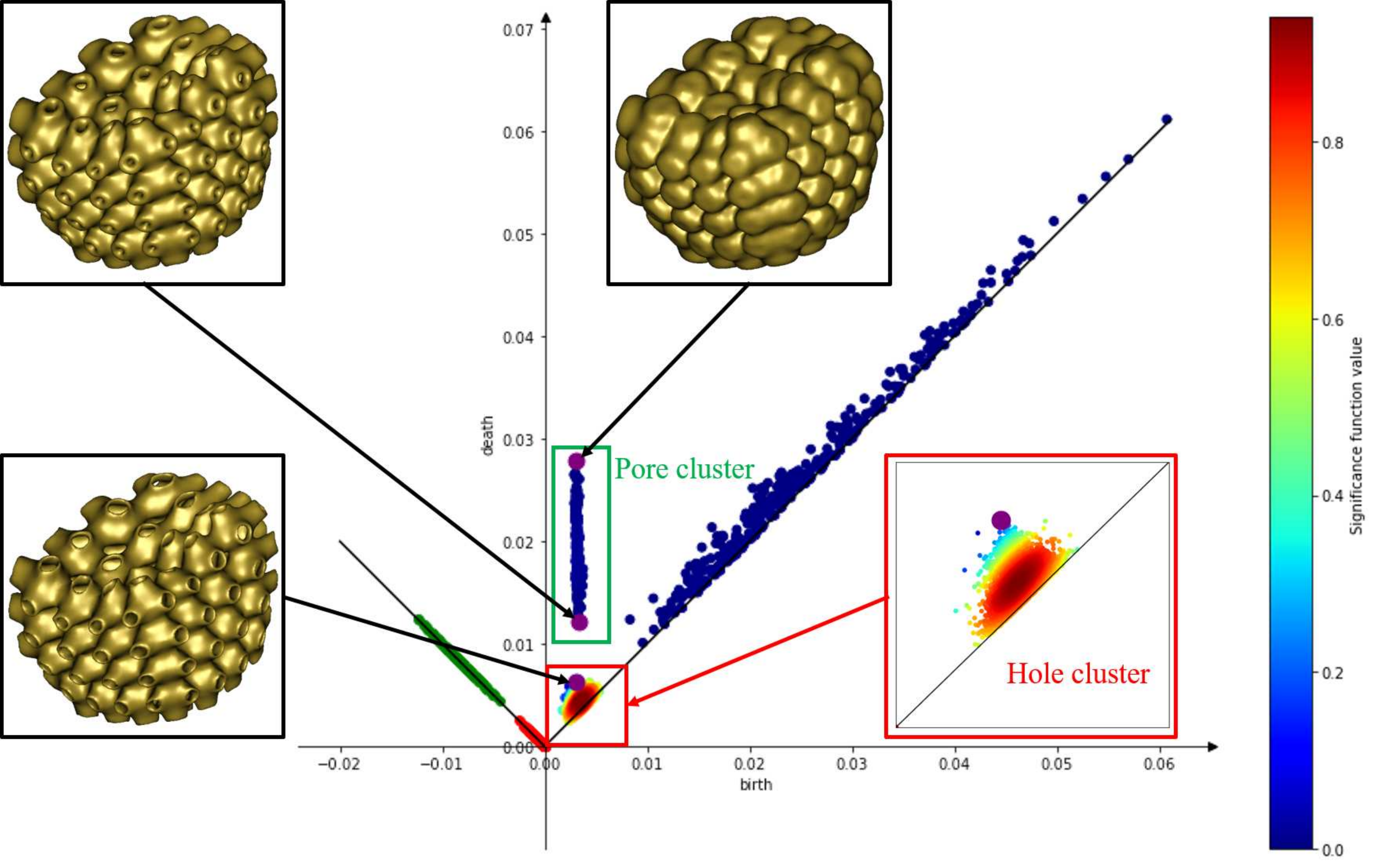}}
    \subfigure[Significance function values of the points in the short persistence cluster.]{\label{subfig:tooth_P_significance} \includegraphics[width=0.45\textwidth]{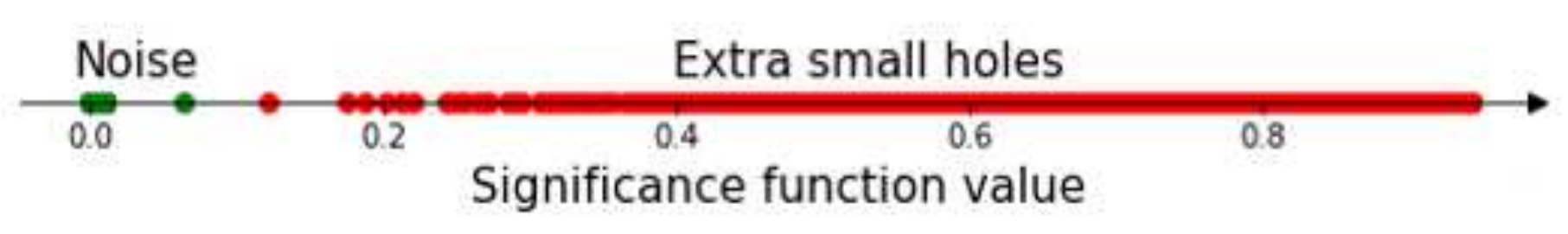}}

    \caption
    {
        Determination of the range of thickness parameter $c$ for the \emph{Tooth P} sheet structure $f(x,y,z) \leq c$ extracted from the scalar field in Fig. \ref{subfig:tooth p scalar}.
        Persistence pairs in 1-PD can be partitioned into three clusters using two rounds of Agglomerative hierarchical clustering algorithms.
    }
    \label{fig:c range_tooth_P}
 \end{figure}

 The pores and extra small holes on the porous sheet structure can be
    regarded as one dimensional cycles (1-cycles) that cannot be shrunk to a pole.
 They are the generators of one dimensional homology groups,
    and can be represented by the points in 1-PD (Fig.~\ref{subfig:tooth_P_PD}) of the distance field $f(x,y,z)$~\pref{eq:implicit}.
 The 1-cycles corresponding to the pores exist much longer time than those to the extra small holes,
    because the sizes of pores are much larger than those of extra small holes
 Each 1-cycle is represented by a point $(x,y)$ in the 1-PD (Fig.~\ref{subfig:tooth_P_PD}),
    where $x$ and $y$ are the birth time and death time, respectively,
    and then $y-x$ indicates the persistence time.
 The persistence time (i.e., $y-x$) of a pore is much larger than that of an extra small hole.
 Thus, in the 1-PD (Fig.~\ref{subfig:tooth_P_PD}),
    the points corresponding to the pores are far away from the line $y=x$,
    whereas those to the extra small holes are much closer to the line $y=x$.
 Therefore, the points in the 1-PD (Fig.~\ref{subfig:tooth_P_PD})
    can be partitioned into two clusters depending on their persistence time,
    and the cluster with longer persistence time corresponds to the pores.
 However, the other cluster with shorter persistence time contains points
    corresponding to extra small holes and topological noises,
    because the persistence times of extra small holes and topological noises are comparable.
 Therefore, we should design a method to perform a distinction between the points
    for extra small holes and those for topological noises
    by studying the properties of extra small holes and topological noises.
 In the generation period of the 1-PD using sub-level filtration
    of the distance field $f(x,y,z)$~\pref{eq:implicit},
    the birth time of extra small holes focuses on the start period,
    same as that of the pores,
    whereas that of topological noises uniformly distributes in the whole 1-PD generation period.
 Thus, the density of the point set to the extra small holes is much larger
    than that of the point set to the topological noises.
 The point set to the extra small holes can be
    distinguished from that to the topological noises given these properties.

 On the above basis of the analysis of the points in the 1-PD,
    we develop a method for determining the point set to the pores
    and to the extra small holes
    using two rounds of clustering.
 As illustrated in Fig.~\ref{subfig:tooth_P_PD},
    in the first round clustering,
    we initially project the points onto the line $y=-x$,
    and the projected points on the line $y=-x$ clearly form two clusters,
    one with longer persistence time
    and the other with shorter persistence time.
 Then, we partition the points into two clusters using the agglomerative hierarchical clustering algorithm~\cite{mullner2011modern}.
 The pores are represented by the points in the cluster with
    longer persistence time, namely, \emph{pore cluster}, and denoted as,
 \begin{equation} \label{eq:pore_cluster}
    \mathcal{C}_p = \{ (b_i^p,d_i^p), i=1,2,\cdots,n_p \},
 \end{equation}
    and the extra small holes and topological noises are represented by the points in the cluster with shorter persistence time,
    namely, \emph{short persistence cluster}, and denoted as,
 \begin{equation} \label{eq:short_cluster}
    \mathcal{C}_{s} = \{(b_i^{s},d_i^{s}\}, i=1,2,\cdots,n_{s} \}.
 \end{equation}
 From the pore cluster, we can generate two thresholds, i.e.,
 \begin{equation} \label{eq:c_max}
    \begin{aligned}
    &c_{max,1} = min\{d_i^p, i=1,2,\cdots,n_p\}, \\
    &c_{max,2} = max\{d_i^p, i=1,2,\cdots,n_p\}.
    \end{aligned}
 \end{equation}
 As previously mentioned, the sheet structure is represented as
    $f(x,y,z) \leq c$.
 When $c < c_{max,1}$, the pores remain open;
 when $c \geq c_{max,1}$, some pores are closed;
 when $c \geq c_{max,2}$, all pores are closed (Fig.~\ref{subfig:tooth_P_PD}).

 In the second round clustering,
    we further segment the short persistence cluster into two sub-clusters,
    one for the extra small holes,
    and the other for the topological noises.
 For this purpose,
    we initially map the points in the short persistence cluster into the interval $[0,1]$ by the \emph{significance function},
    i.e.,
 \begin{equation} \label{eq:sig_function}
    {g}(x,y)=(1-\frac{x}{{{b}_{max }}}) \frac{ t(x,y)}{t_{max}},
 \end{equation}
 where $$b_{max} = \max\{b_i^s, i=1,2,\cdots,n_s\},$$
 $$ t_{max} = \max\{t(b_i^s,d_i^s), i=1,2,\cdots,n_s\},$$
 and
 \begin{equation}
    t(x,y) = \frac{1}{n_s h} \sum_{i=1}^{n_s} exp\left(\frac{(x-b_i^s)^2+(y-d_i^s)^2}{h}\right).
 \end{equation}
 Here, $h=n_s^{-\frac{1}{dim+4}}$ is the bandwidth designated by the
    Scott's rule~\cite{scott2015multivariate},
    and $dim$ is the dimension of the space the points lie in.
 In our case, $dim = 2$.
 The significance function measures the significance of the points in the
    short persistence cluster by considering the birth time and the point density.
 The topological noises uniformly appear and distribute in the period
    of 1-PD generation,
    which assigns a small value to the point corresponding to topological noises.
 However, because the birth times of extra small holes focus on the early
    stage of the 1-PD generation period,
    it provides a large value to the point corresponding to the extra small hole.
 Therefore, as illustrated in Fig.~\ref{subfig:tooth_P_significance},
    the significance function~\pref{eq:sig_function} maps the points corresponding to topological noises to the left part of the interval $[0,1]$,
    whereas it maps the points corresponding to extra small holes to the right part of the interval $[0,1]$.
 The points mapped in the interval $[0,1]$ can be segmented into
    two clusters by the agglomerative hierarchical clustering algorithm,
    and accordingly, the short persistence cluster is partitioned into
    two sub-clusters,
    one for topological noises,
    and the other for extra small holes.
 We refer to the sub-cluster for extra small holes as \emph{hole cluster},
    and denote it as,
 \begin{equation} \label{eq:hole_cluster}
    \mathcal{C}_h = \{(b_i^h, d_i^h),\ i=1,2,\cdots,n_h\},
 \end{equation}
    and we refer to the sub-cluster for topological noises as \emph{topological noise cluster},
    and denote it as,
 \begin{equation} \label{eq:noise_cluster}
    \mathcal{C}_t = \{(b_i^t, d_i^t),\ i=1,2,\cdots,n_t\}.
 \end{equation}

 With the hole cluster, we can obtain another threshold, i.e.,
 \begin{equation} \label{eq:c_min}
    c_{min} = \max\{d_i^h,\ i=1,2,\cdots,n_h\}.
 \end{equation}
 When $c \geq c_{min}$, the extra small holes disappear from the sheet structure
    $f(x,y,z) \leq c$.

\subsection{Topology control in porous sheet generation}
\label{subsec:selection}

 We can control the topology structure of the generated porous sheet structure by using the three thresholds,
    i.e.,
    $c_{max,1}$,
    $c_{max,2}$~\pref{eq:c_max},
    and $c_{min}$~\pref{eq:c_min}.
 The death times of the points in the
    pore cluster~\pref{eq:pore_cluster} is assumed to be ordered, i.e.,
    $$c_{max,1}=d_1^p < d_2^p < \cdots < d_{n_p}^p = c_{max,2}.$$
 Moreover, because PD is a multi-set, i.e.,
    several points are probably at one position,
    and $k_i$ points are assumed to be at the position $(b_i^p,d_i^p),\ i=1,2,\cdots,n_p$.

 In general, the porous sheet structure can be generated by the following
    optimization problem:
 \begin{align}
    &\min_c \ B(c) \nonumber\\
    &s.t. \ c\in[{{c}_{min }},c_e), \nonumber\
    \label{eq:op}
 \end{align}
    where $B(c)$ is the objective function,
    and $c$ is the thickness parameter.
 The objective function $B(c)$ can be constructed using,
    such as compliance, porosity, etc.
 To demonstrate the topology control capability,
    we generate the pore sheet structure by setting
    $$ B(c) =  \left\| V(c)-{{V}_{0}} \right\|,$$
 where $V(c)$ is the volume ratio of the porous sheet structure
    $f(x,y,z) \leq c$ to its AABB,
    and $V_0$ is a given value.
 If $c_{min} \leq c_e < c_{max,1}$,
    the generated sheet structure has no extra small holes,
    and all the pores remain open.
 When $ c_{max,1} \leq c_e < c_{max,2}$,
    some pores are probably closed;
    when $c_e \geq c_{max,2}$,
    all pores are probably closed.
 More precisely, if $c_e = d_i^p, i=1,2,\cdots,n_p$,
    at most $\sum_{j=1}^i k_j$ pores will probably be closed.

\section{Experimental results and discussion}
\label{sec:experimental}

 In this section,
     some experiments have been conducted to evaluate the performance of the proposed algorithm.
 All experiments are implemented in Python on a personal computer with
     Intel(R) Core (TM) i7-7700 CPU @ 3.6 GHz and 16 GB RAM.
 Some open source libraries are used in this study.
 The Taichi library \cite{hu2019taichi} is used for multi-threading to
    fit DUDF based on LSPIA,
    and the GUDHI Python module \cite{maria2014gudhi} is adopted to generate cubical complex and calculate PDs.
 In our implementation, we sample point clouds from four types of typical TPMS \cite{von1991nodal,rajagopalan2006schwarz} (Schwarz's Diamond (\emph{D}) surface, Schoen's Gyroid (\emph{G}) surface, \emph{I-WP} surface, and Schwarz's Primitive (\emph{P}) surface),
    as well as some porous models generated using the method
    in Ref.~\cite{hu2021heterogeneous}.

 \begin{table*}[!htb]
    \centering
    \caption{The performance of fitting the DUDF generated from the input point clouds.}
    \label{table:fit}
    \resizebox{0.75\textwidth}{!}{
    \begin{threeparttable}
    \begin{tabular}{ccccccc}
    \hline
    Model           & Fig.                           & Data points  & Grid size\tnote{1}        & Iteration   & Average Error   & Time (seconds) \\ \hline
    D               & \ref{subfig:sheet_D_PD}        & 73,481       & $128\times128\times128$   & 44          & $1.92 \times {10}^{-5}$          & 20.95          \\
    G               & \ref{subfig:sheet_G_PD}        & 73,176       & $128\times128\times128$   & 41          & $1.83 \times {10}^{-5}$             & 19.92          \\
    I-WP            & \ref{fig:I-WP unit surface}    & 10,405       & $64\times64\times64$      & 50          & $8.05 \times {10}^{-4}$              & 24.23          \\
    P               & \ref{subfig:sheet_P_PD}        & 73,601       & $128\times128\times128$   & 32          & $1.17 \times {10}^{-5}$              & 18.32          \\
    Tooth P         & \ref{fig:Illustration}         & 71,933       & $128\times128\times128$   & 35          & $8.39 \times {10}^{-6}$              & 18.87          \\
    Venus I-WP      & \ref{subfig:venus_IWP_32}      & 68,865       & $32\times32\times32$      & 42          & $2.71 \times {10}^{-3}$              & 19.67          \\
    Venus I-WP      & \ref{subfig:venus_IWP_64}      & 68,865       & $64\times64\times64$      & 42          & $1.25 \times {10}^{-3}$              & 21.05          \\
    Venus I-WP      & \ref{subfig:venus_IWP_128}     & 68,865       & $128\times128\times128$   & 42          & $8.44 \times {10}^{-6}$           & 20.61          \\
    Balljoint I-WP  & \ref{subfig:high_density_fit}  & 14,643       & $64\times64\times64$      & 46          & $1.47 \times {10}^{-3}$              & 22.49          \\ \hline
    \end{tabular}
    \begin{tablenotes}
    \footnotesize
    \item[1] The size of the control grid of the trivariate B-spline function.
    \end{tablenotes}
    \end{threeparttable}
    }
 \end{table*}

\subsection{Efficiency, Robustness and Comparison}
\label{subsec:Eff sheet}

 \begin{figure*}[!htb]
    \centering	
    \subfigure[]{\includegraphics[width=0.14\textwidth]{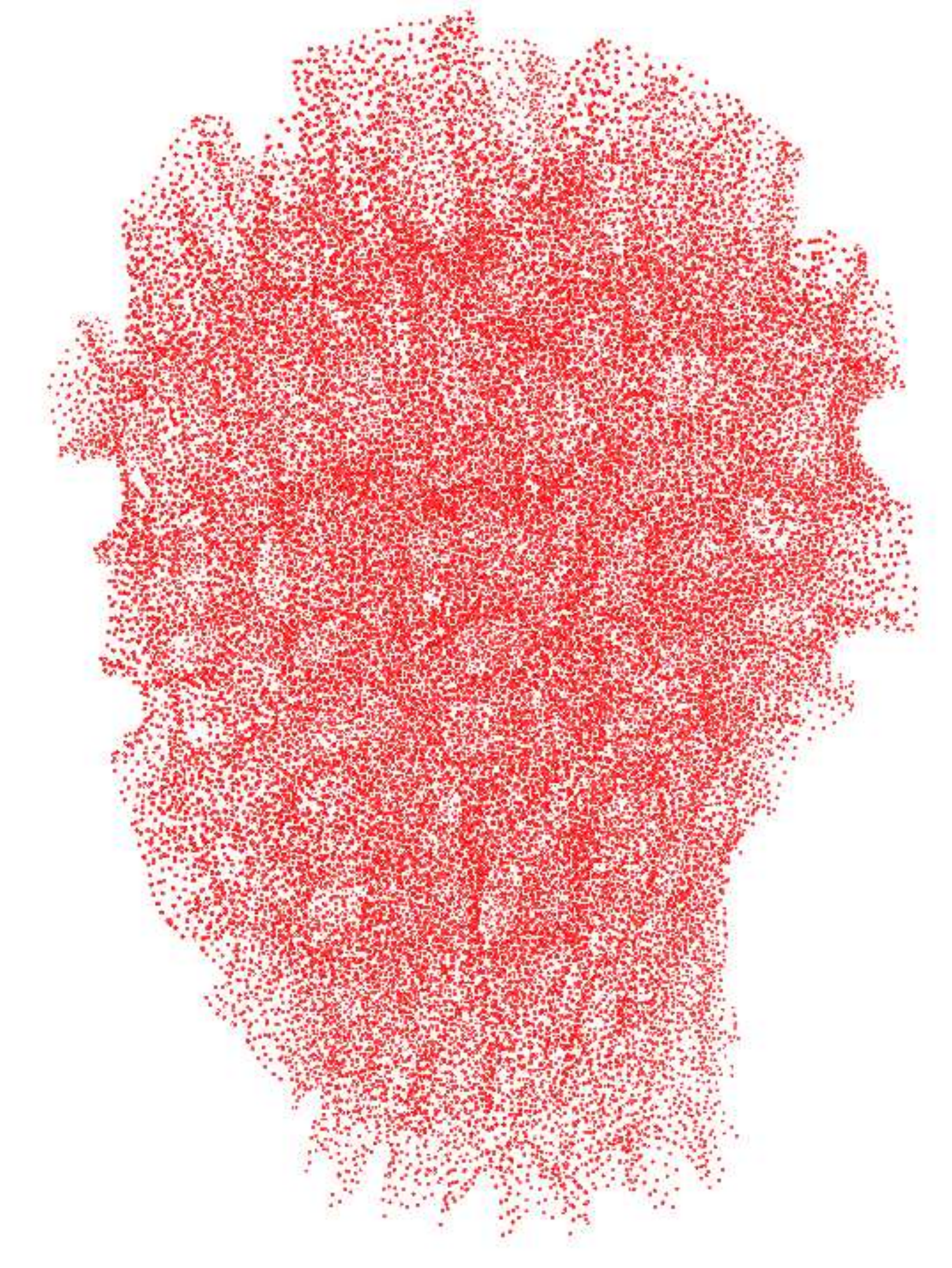}}
    \quad
    \subfigure[]{\label{subfig:venus_IWP_32} \includegraphics[width=0.14\textwidth]{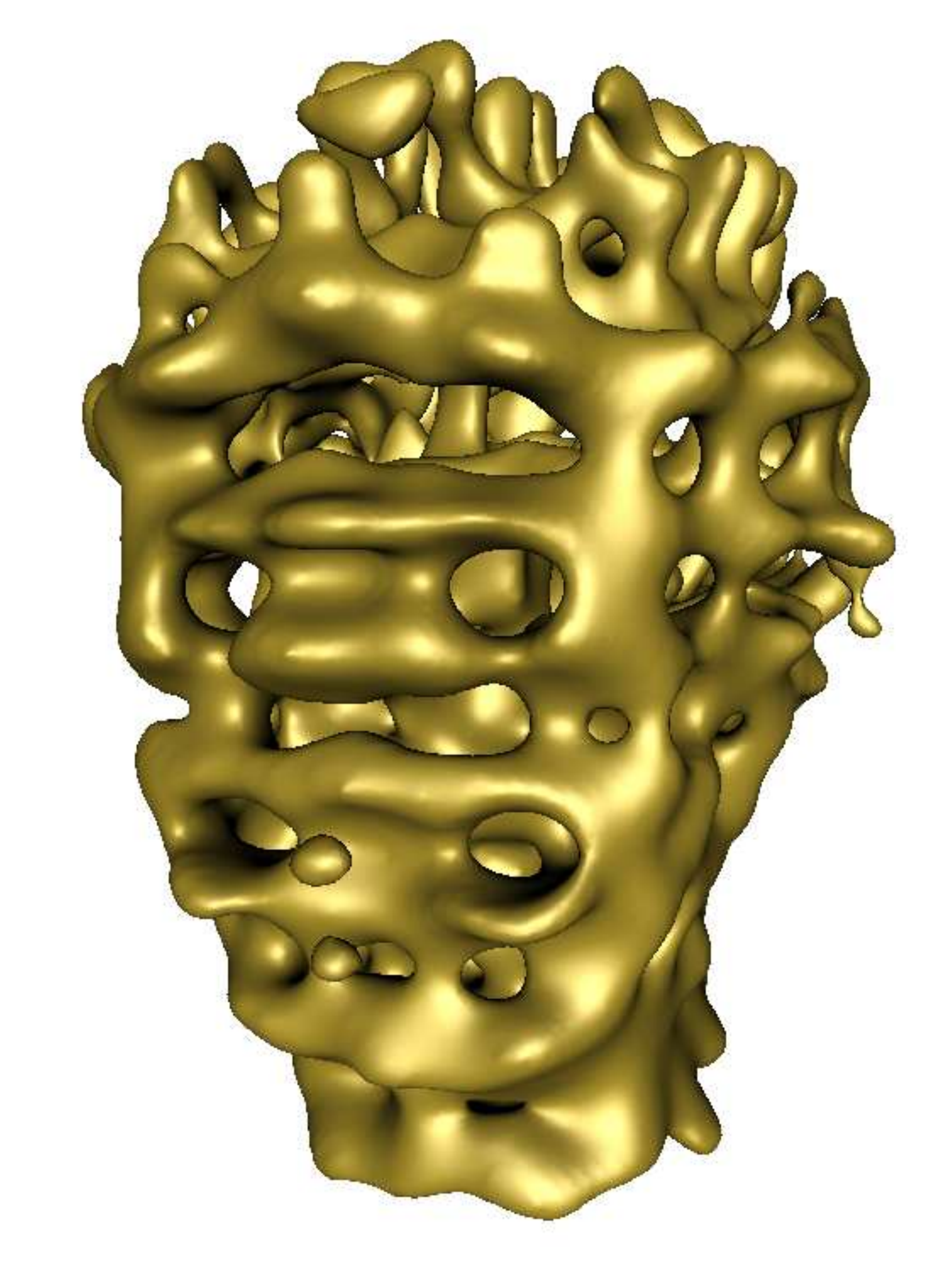}}
    \quad
    \subfigure[]{\label{subfig:venus_IWP_64} \includegraphics[width=0.14\textwidth]{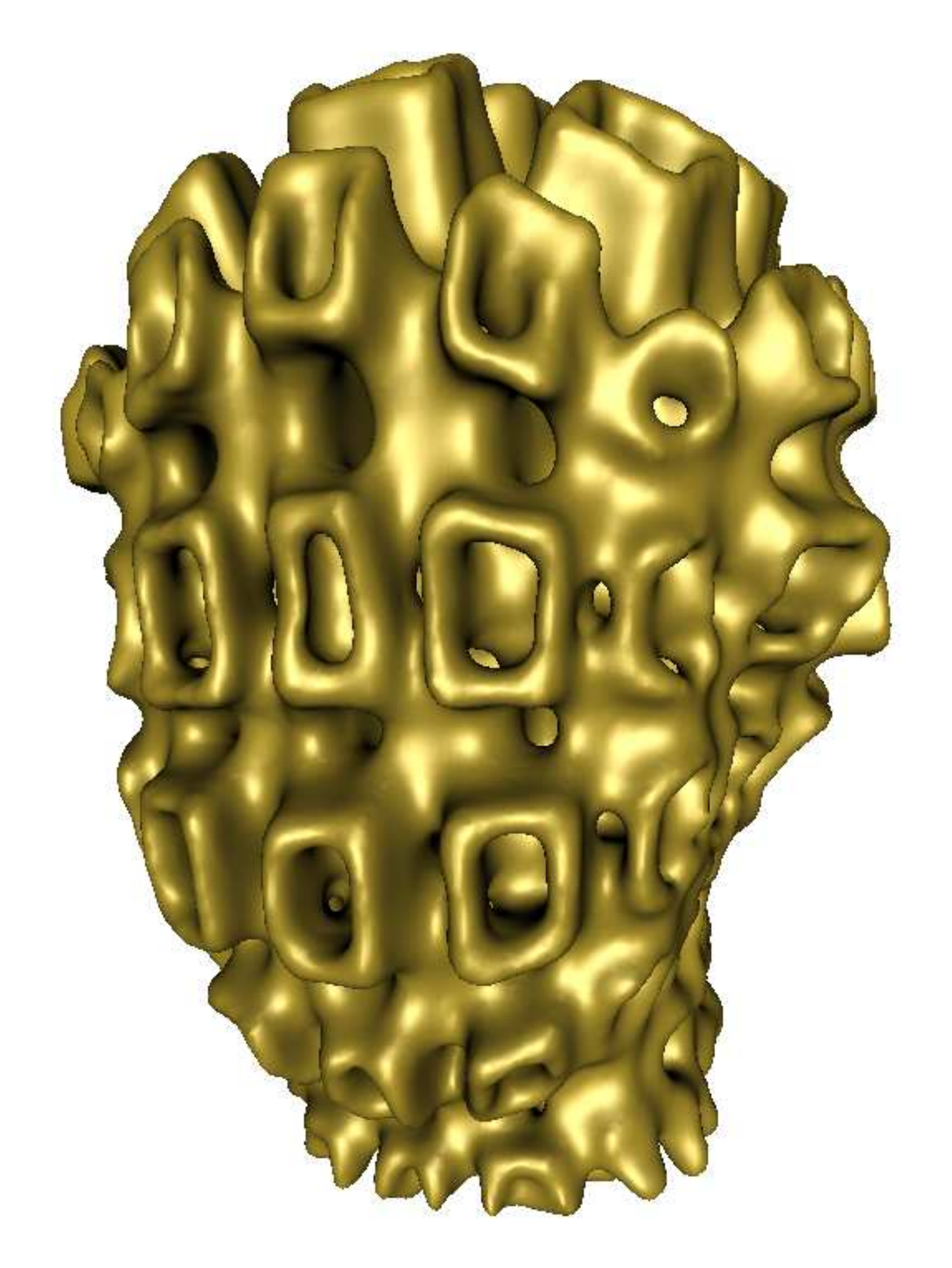}}
    \quad
    \subfigure[]{\label{subfig:venus_IWP_128} \includegraphics[width=0.14\textwidth]{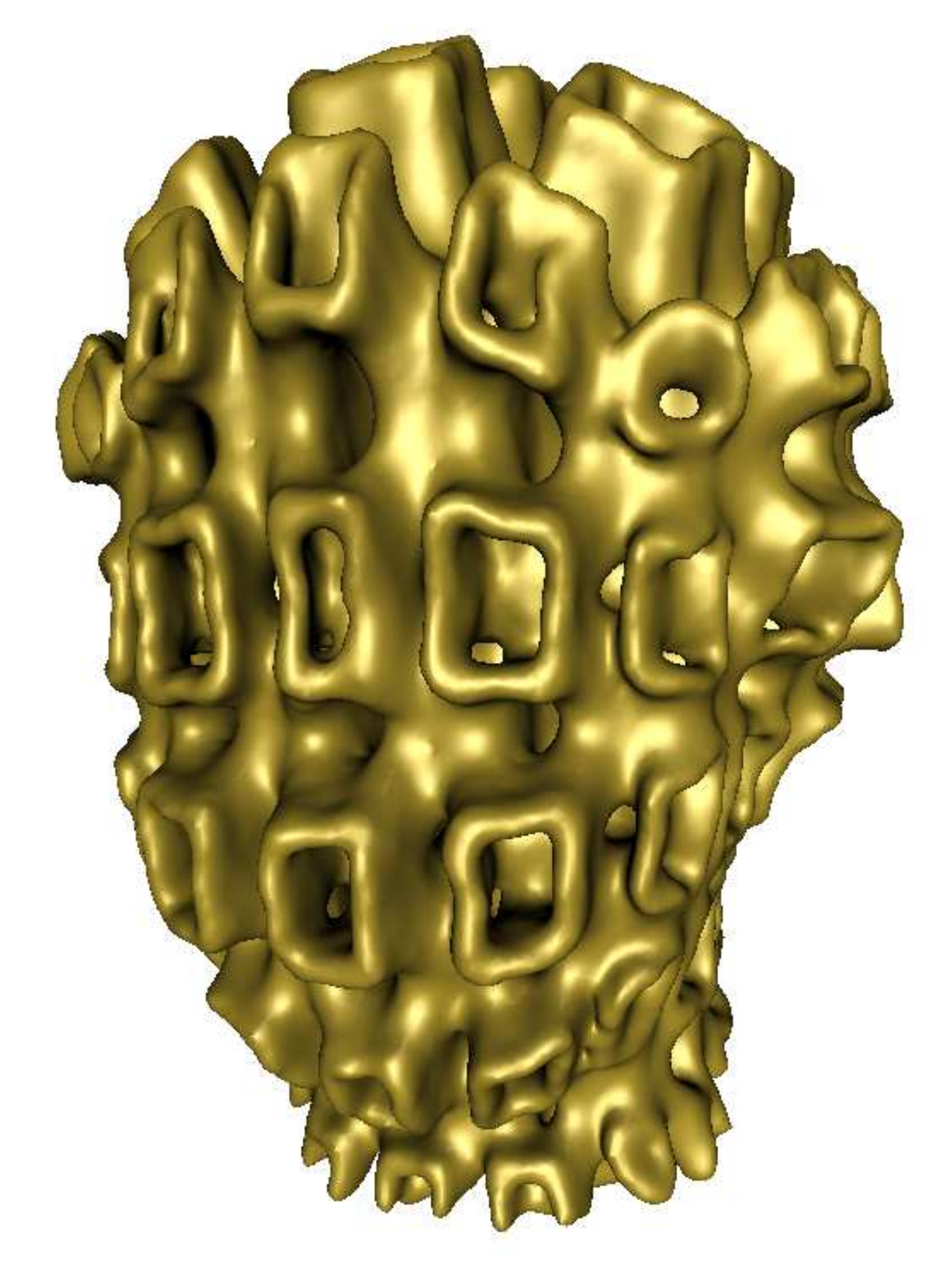}}

    \caption
    {
        Point cloud (\emph{Venus I-WP}) with 68,865 points (a),
            and porous sheet structures $f(x,y,z) \leq 0.008$ with the control grid size of $32 \times 32 \times 32$ (b), $64 \times 64 \times 64$ (c) and $128 \times 128 \times 128$ (d).
        Details become clear as the control grid size increases.
    }
    \label{fig:venus_porous_IWP surface}
 \end{figure*}
 \begin{figure*}[!htb]
    \centering	
    \subfigure[]{\label{subfig:high density} \includegraphics[width=0.135\textwidth]{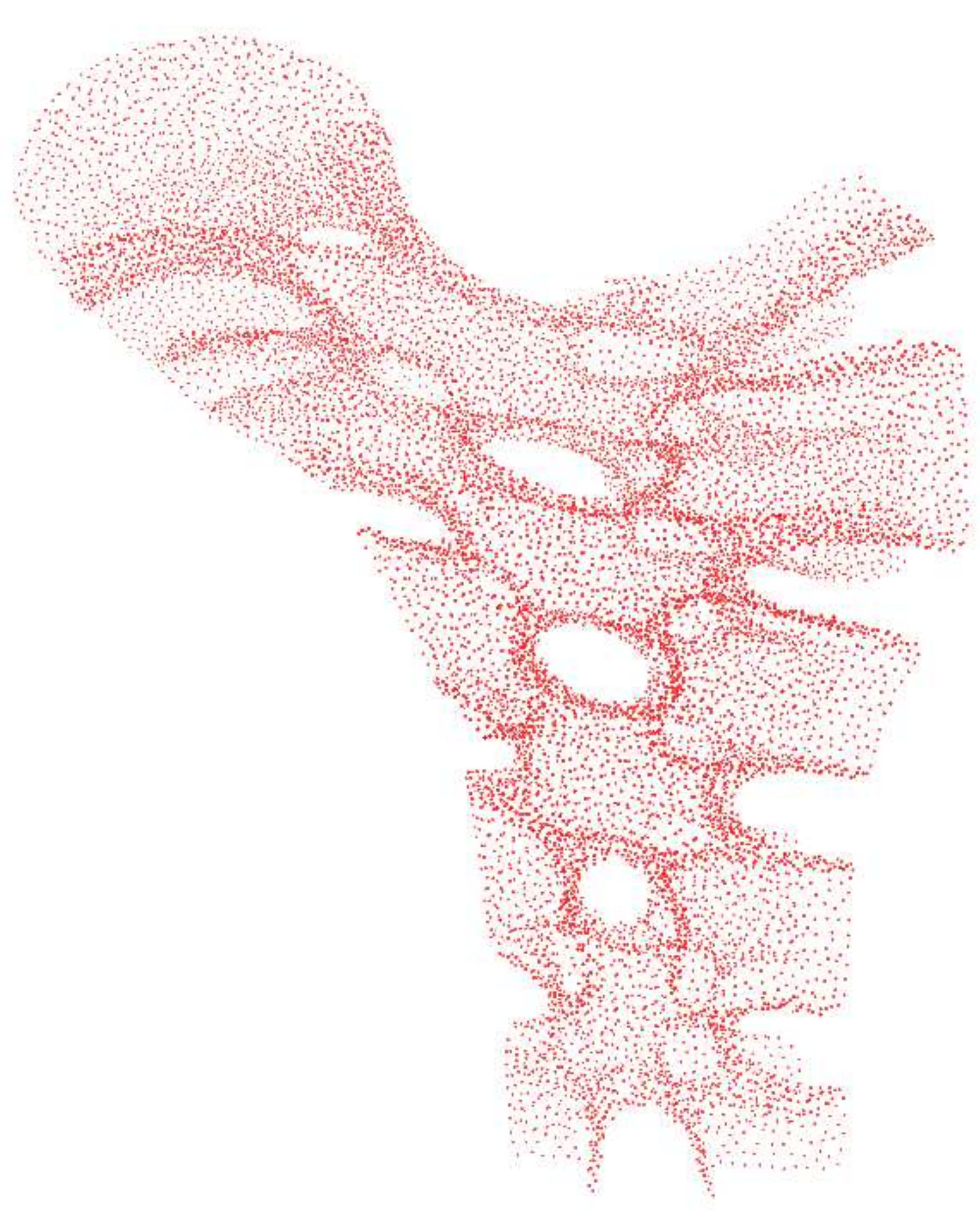}}
    \quad
    \subfigure[]{\label{subfig:high_density_fit} \includegraphics[width=0.135\textwidth]{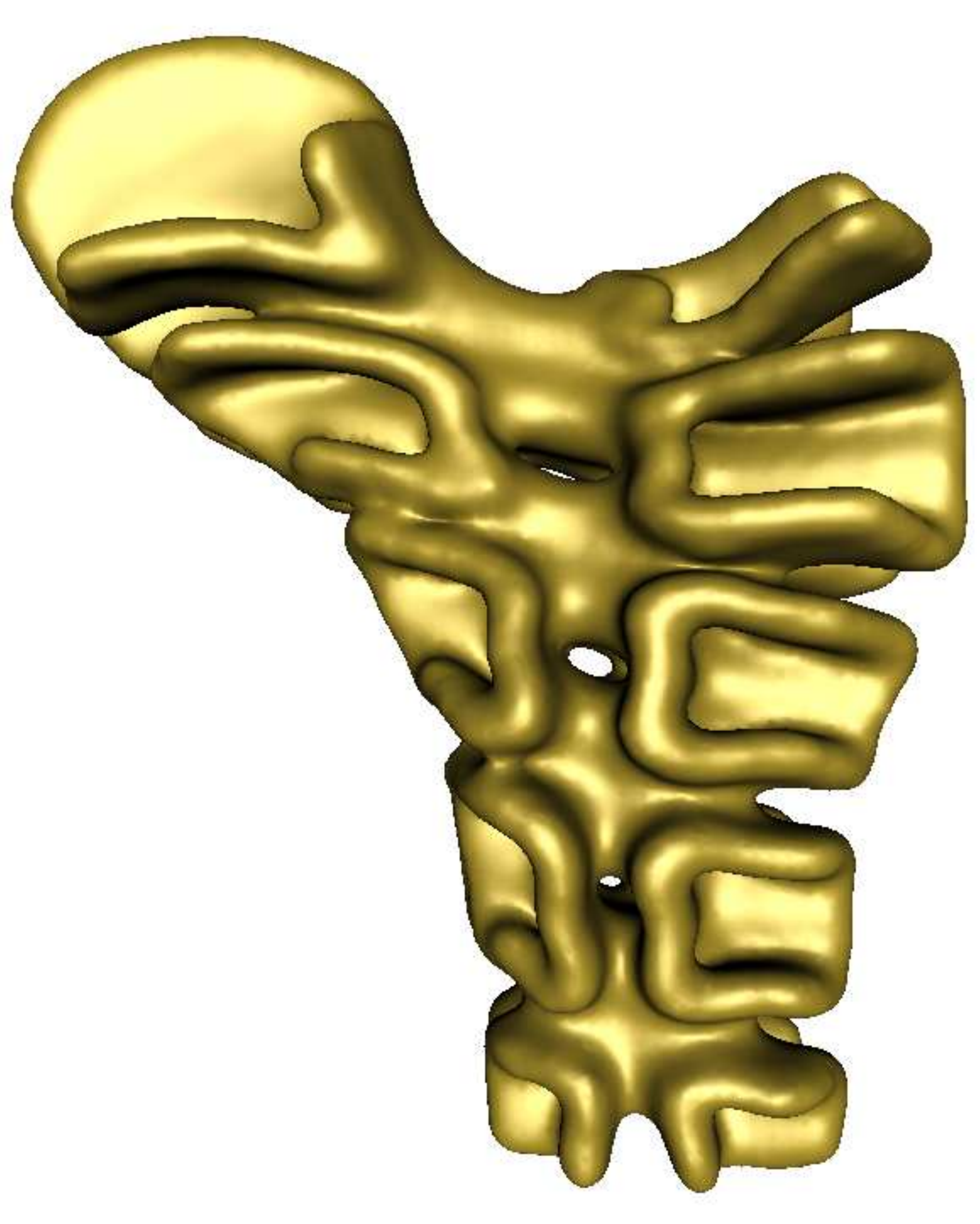}}
    \quad
    \subfigure[]{\label{subfig:low density} \includegraphics[width=0.135\textwidth]{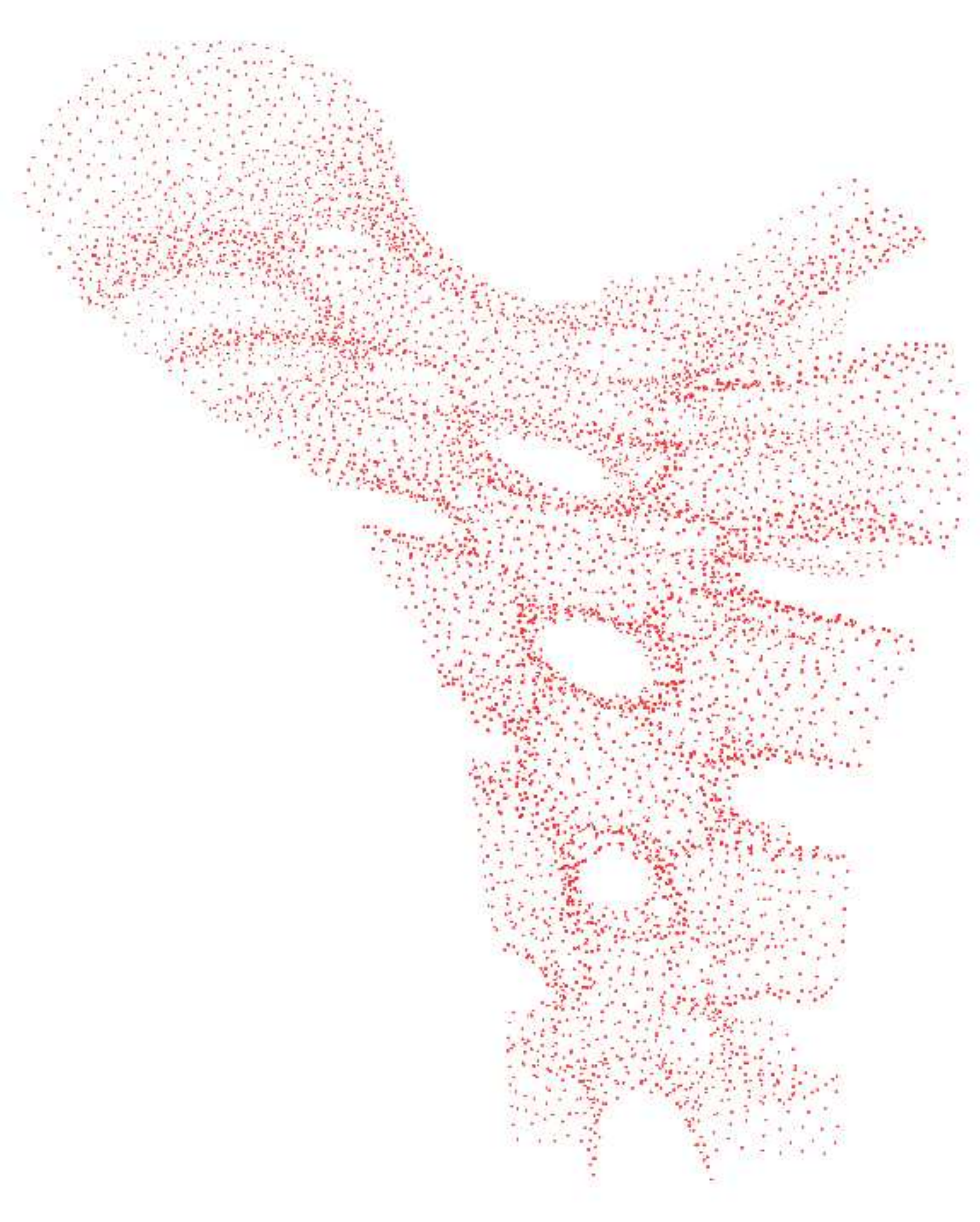}}
    \quad	
    \subfigure[]{\includegraphics[width=0.135\textwidth]{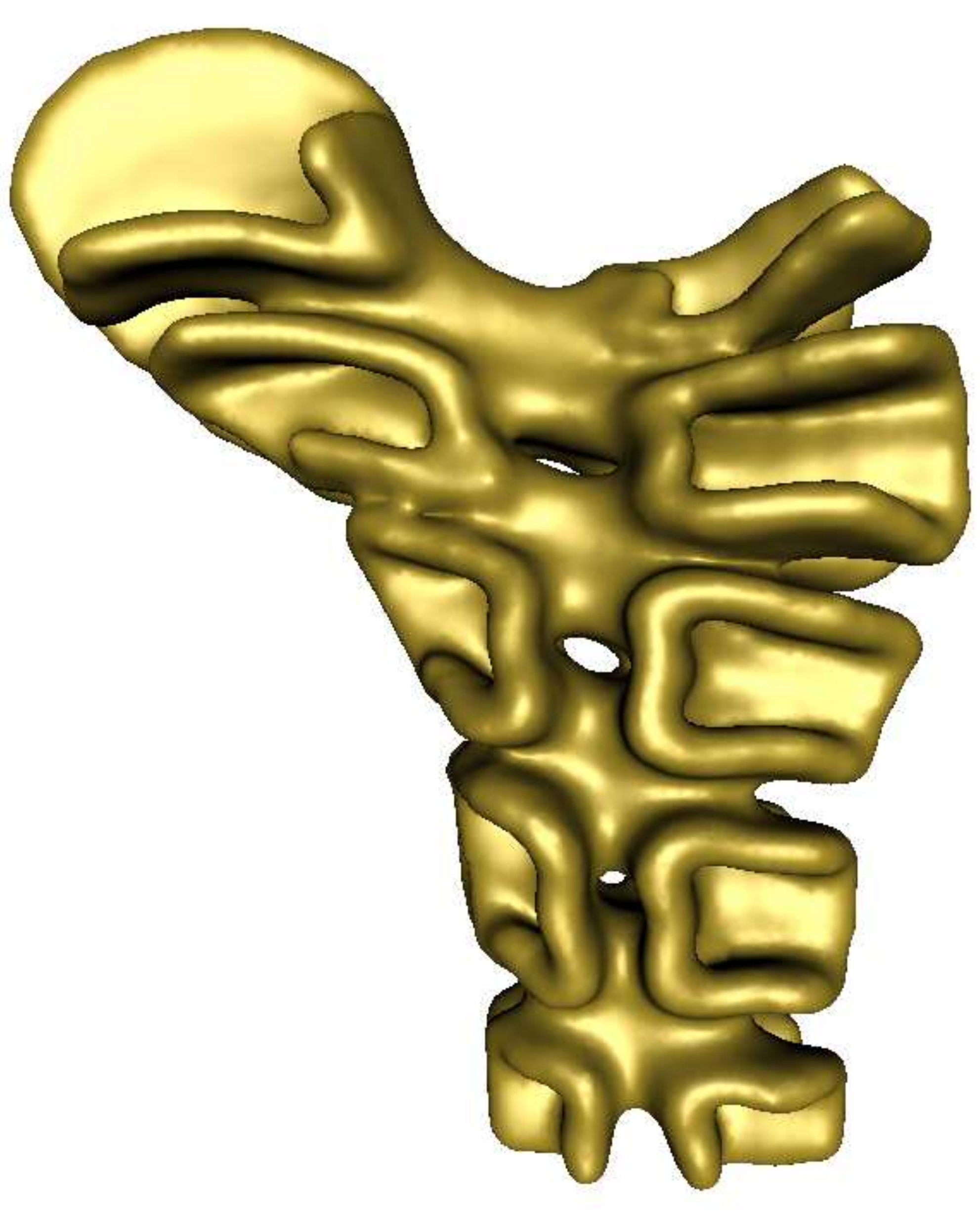}}

    \caption
    {
        The sheet structures generated from point clouds with different densities,
            where the Hausdorff distance of two sheet structures (b, d) with respect to the diagonal of the AABB is $1.15 \times {10}^{-2}$.
        (a) High density point cloud (\emph{Balljoint I-WP}) with 14643 points.
        (b) Porous sheet structure $f(x,y,z) \leq 0.015$ with a $64 \times 64 \times 64$ control grid generated from point cloud in (a).
        (c) Low density point cloud (\emph{Balljoint I-WP}) with 6,579 points.
        (d) Porous sheet structure $f(x,y,z) \leq 0.015$ with a $64 \times 64 \times 64$ control grid generated from point cloud in (c).
    }
    \label{fig:Balljoint_porous surface}
 \end{figure*}

 In this section,
    several porous sheet structures with implicit B-spline representations
    are directly generated by fitting the point clouds using the developed method.
 The experimental data for fitting different porous point cloud models
    are listed in Table \ref{table:fit},
    and the average fitting error is
    \[err=\frac{1}{m}\sum\limits_{k=1}^{m}{\frac{\left\| {{d}_{k}}-f({{\boldsymbol{g}}_{k}}) \right\|}{L}},\]
    where $f(\cdot)$ is the fitting trivariate B-spline function,
    ${{Dist}_{\mathcal{P}}}= \{({{\boldsymbol{g}}_{i}},{{d}_{i}}),i=1,2,\cdots,m\}$ is the DUDF,
    and $L$ is the diagonal length of the AABB of the input point cloud.

 Initially, the DUDF fitting precision using LSPIA is affected by the size
    of the control grid of the trivariate B-spline function~\pref{eq:implicit}.
 As shown in Fig.~\ref{fig:venus_porous_IWP surface},
    we fit the point cloud \emph{Venus I-WP} by trivariate B-spline functions with three control grids of different sizes ($32\times32\times32$, $64\times64\times64$ and $128\times128\times128$).
 We can improve the fitting precision by increasing the control grid size, (refer to Table~\ref{table:fit}).
 As shown in Fig.~\ref{fig:venus_porous_IWP surface},
    when the size of the control grid increases,
    the details and features of the resulting implicit B-spline surface become clear.
 With the parallelization of the algorithm,
    LSPIA converges in a short time.
 In all the examples,
    the running time consumed by LSPIA is less than $25$ seconds (refer to Table~\ref{table:fit}).

 Secondly, the DUDF fitting is robust for point clouds with different densities.
 As shown in Fig.~\ref{fig:Balljoint_porous surface},
    the density of the point cloud in Fig.~\ref{subfig:low density} is $45\%$ of the density of the point cloud in Fig.~\ref{subfig:high density}.
 We fit the two point clouds
    by using the trivariate B-spline function with a $64 \times 64 \times 64$ control grid (Fig.~\ref{fig:Balljoint_porous surface}).
 The Hausdorff distance of the two generated sheet structures with respect to the diagonal length of the AABB is $1.15 \times {10}^{-2}$.
 The generated sheet structures shown in Fig.~\ref{fig:Balljoint_porous surface} are nearly the same.

 \begin{figure}[!htb]
    \centering	
    \subfigure[]{\label{subfig:noise} \includegraphics[width=0.135\textwidth]{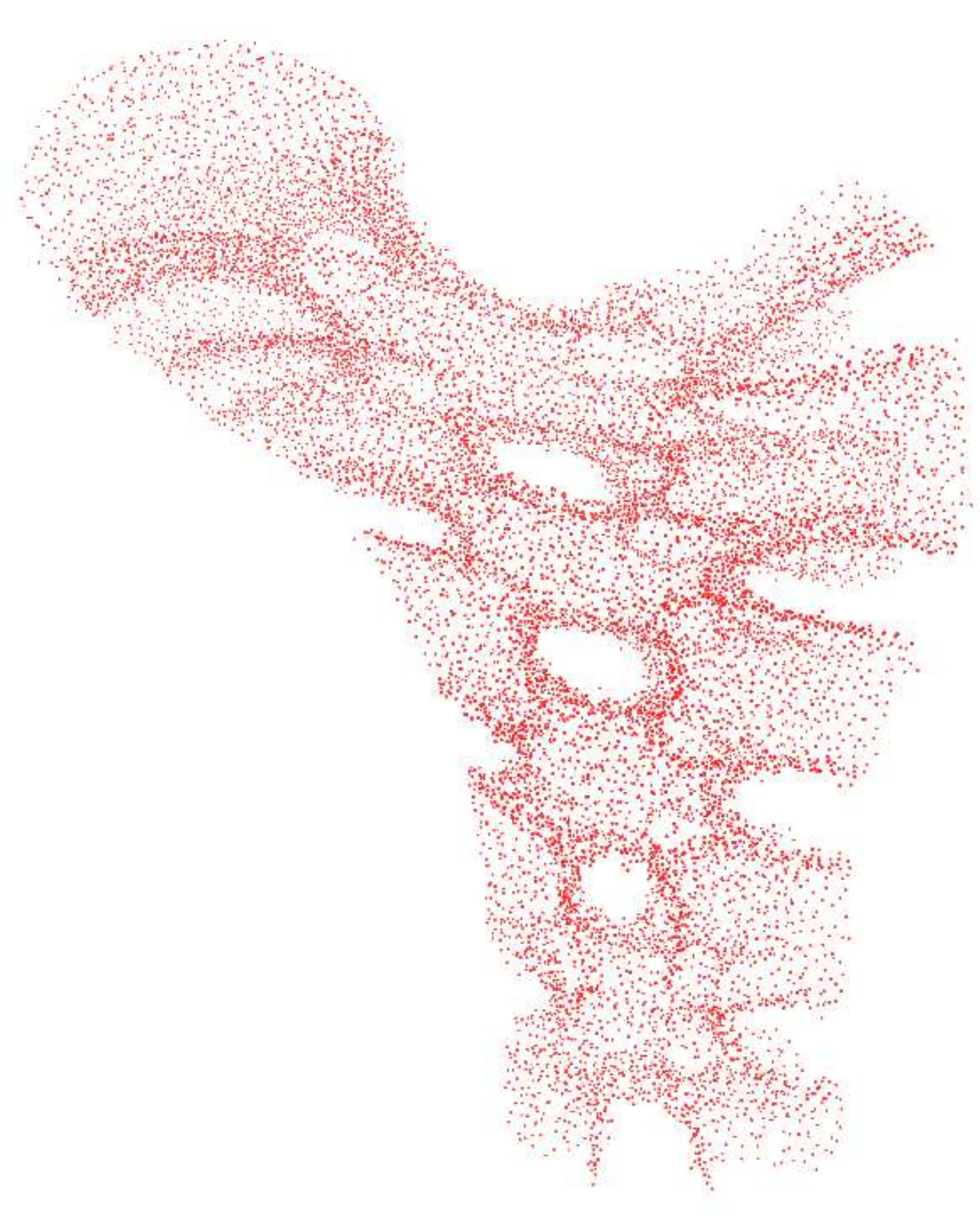}}
    \qquad
    \subfigure[]{\label{subfig:noise_fit} \includegraphics[width=0.135\textwidth]{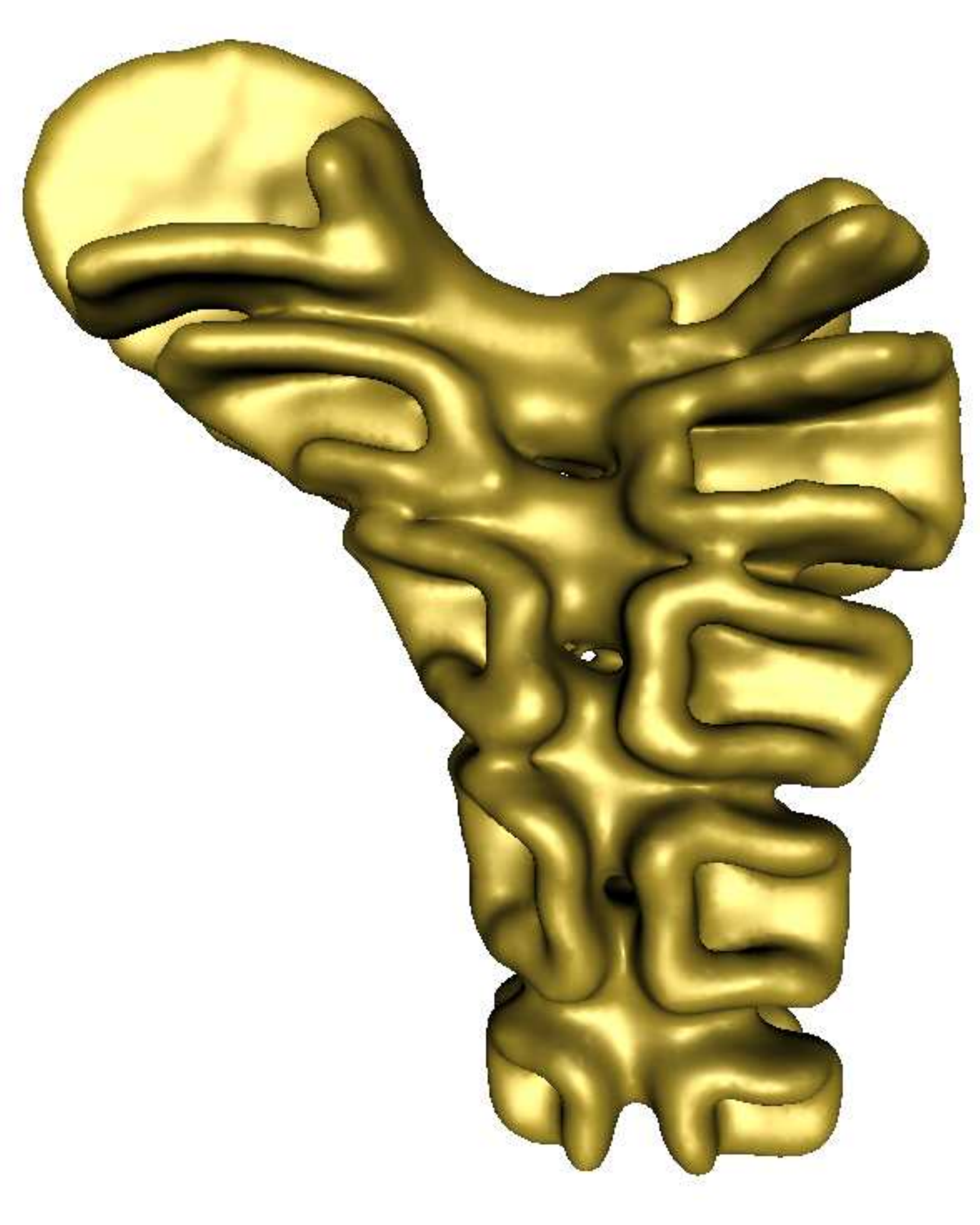}}

    \caption
    {
        The sheet structures generated from the noisy point cloud,
            where the Hausdroff distance of two sheet structures in (b) and Fig.~\ref{subfig:high_density_fit} with respect to the diagonal of the AABB is $2.35 \times {10}^{-2}$.
        (a) Noisy point cloud (\emph{Balljoint I-WP}) generated by adding random noise to the initial point cloud in Fig.~\ref{subfig:high density}.
        (b) Sheet structure $f(x,y,z) \leq 0.015$ with a $64 \times 64 \times 64$ control grid size generated from noisy point cloud in (a).
    }
    \label{fig:Balljoint_porous surface_noise}
 \end{figure}
 Moreover, the DUDF fitting is robust to noise in point cloud.
 As shown in Fig.~\ref{subfig:noise},
    the noisy point cloud is generated by adding random noise to the initial point cloud in Fig.~\ref{subfig:high density}.
 The difference between the generated sheet structures with the same control grid size of $64 \times 64 \times 64$ for the initial point cloud and the noisy point cloud is small,
    as shown in Figs. \ref{subfig:high_density_fit} and \ref{subfig:noise_fit}.
 The Hausdorff distance of the two sheet structures considering the diagonal length of the AABB is $2.35 \times {10}^{-2}$.

\begin{figure}[!htb]
    \centering
    \subfigure[]{\includegraphics[width=0.15\textwidth]{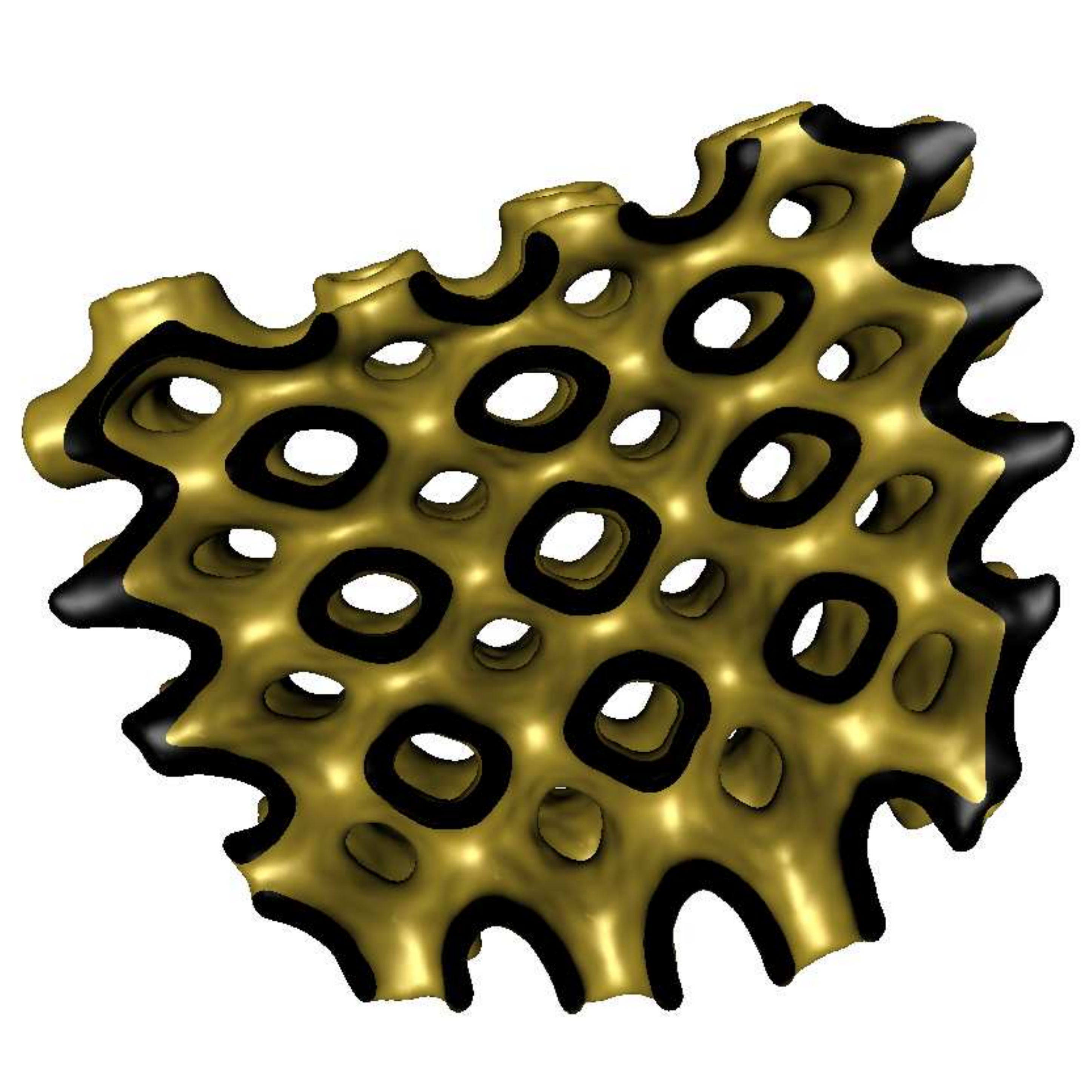}}
    \quad
    \subfigure[]{\includegraphics[width=0.15\textwidth]{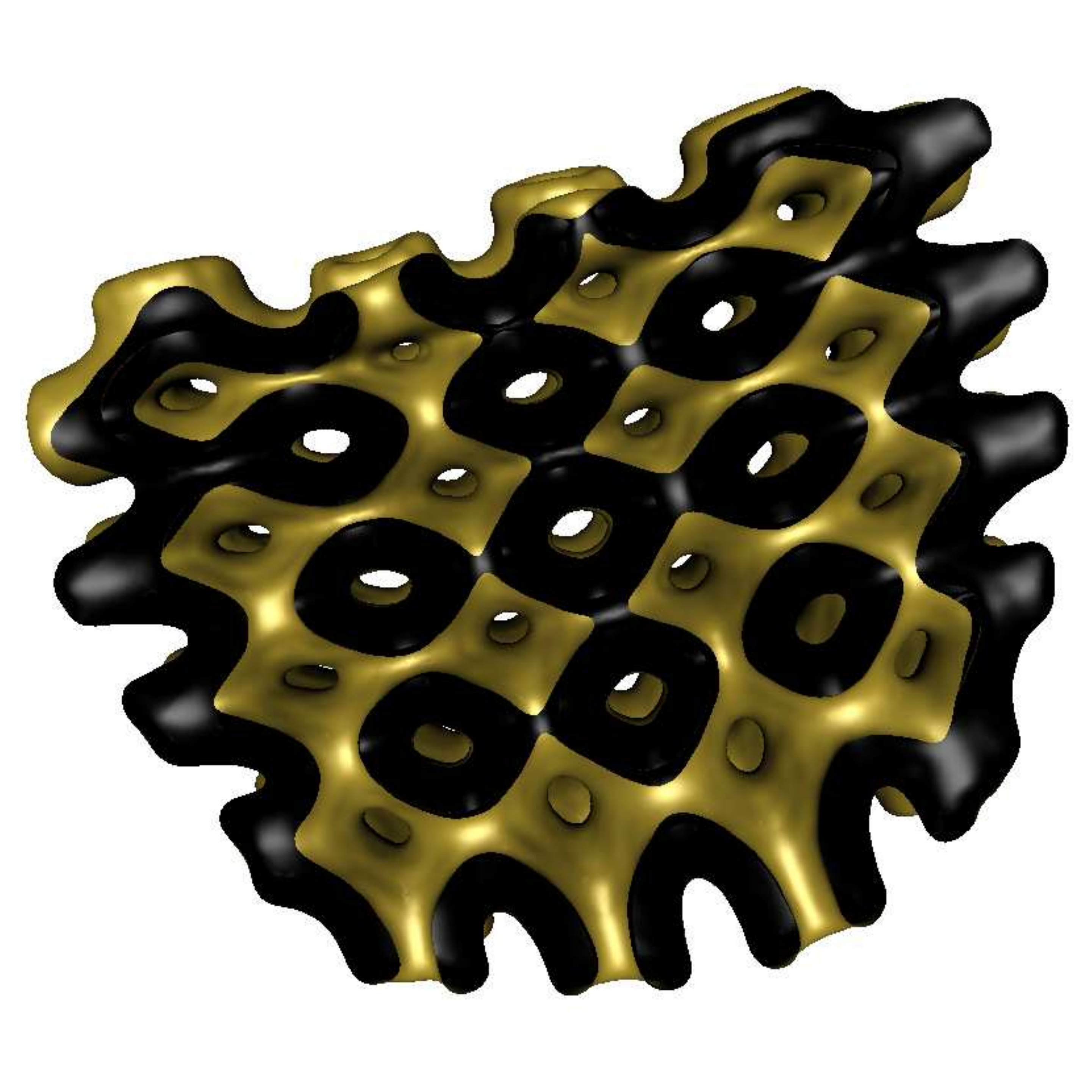}}

    \subfigure[]{\includegraphics[width=0.15\textwidth]{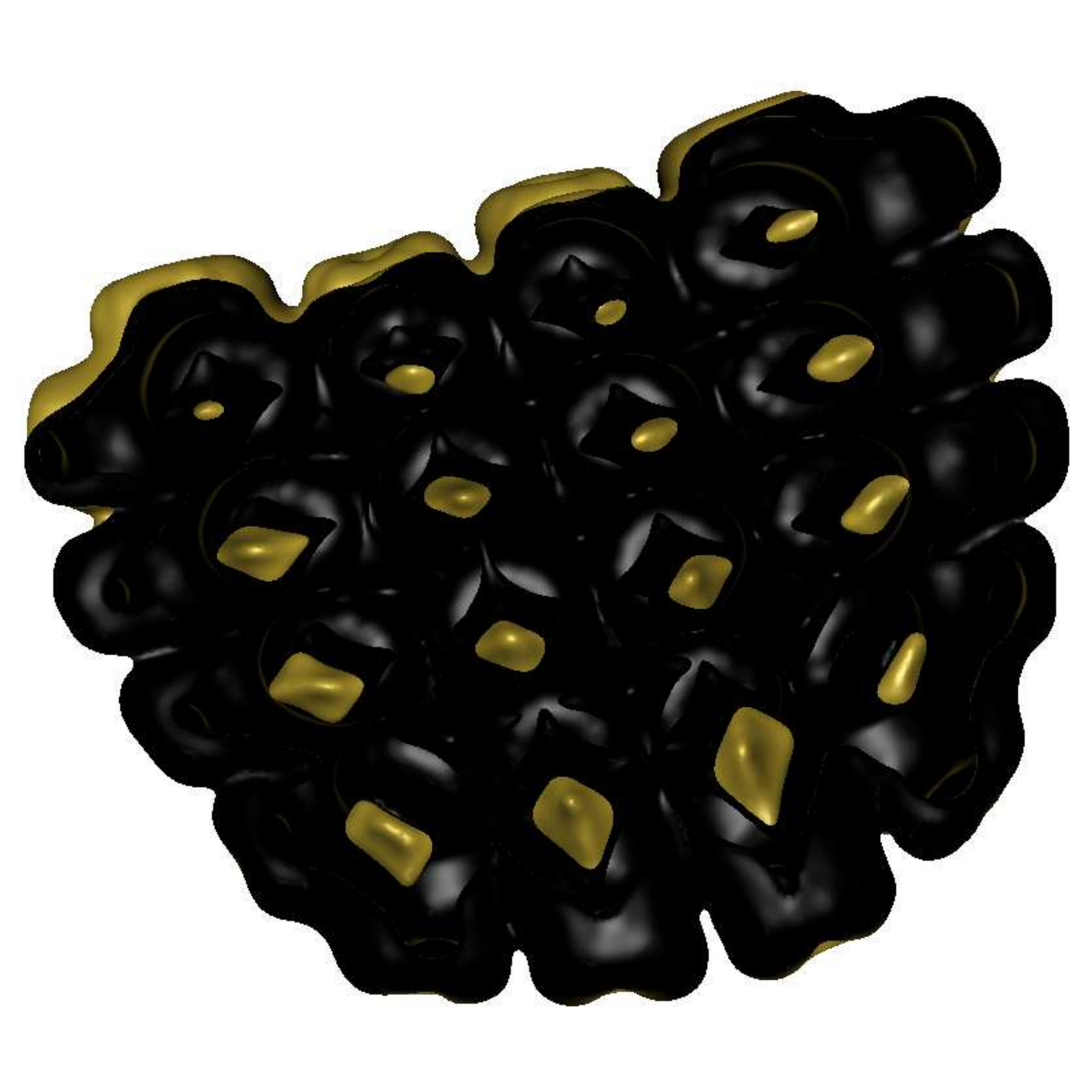}}
    \quad
    \subfigure[]{\includegraphics[width=0.15\textwidth]{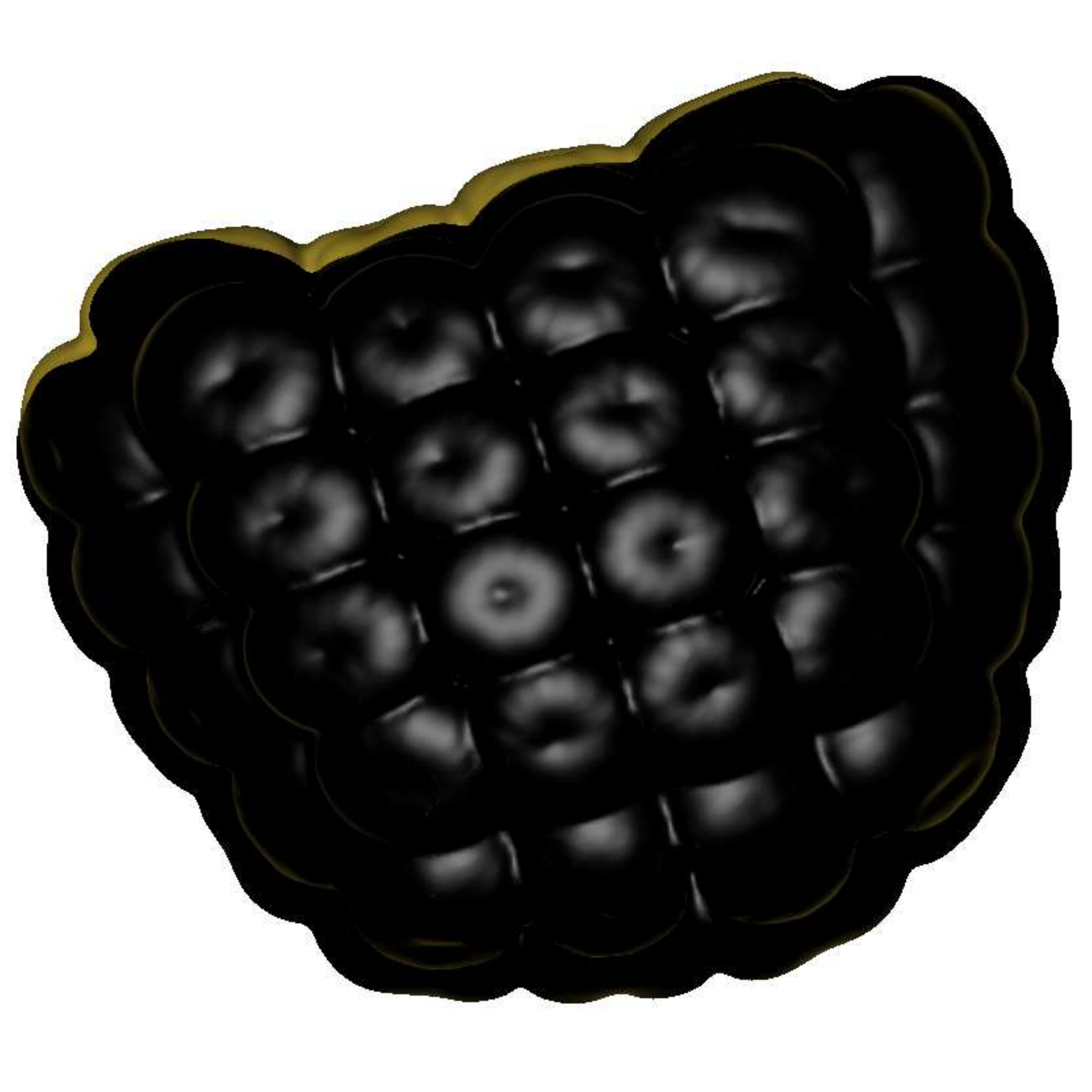}}

    \caption
    {
        The cutaway view of the \emph{Tooth P} sheet structures $f(x,y,z) \leq c$ with different thickness parameter $c$,
            where there is no self-intersection.
            (a) $c=0.0075$. (b) $c=0.015$. (c) $c=0.03$. (d) $c=0.06$.
    }
    \label{fig:c range}
 \end{figure}

 In addition, because the porous sheet structure is represented by
    implicit B-spline function,
    the implicit porous sheet structure  $f(x,y,z) \leq c$ is naturally closed,
    and the internal region has no self-intersection
    for any given thickness parameter $c$,
    as shown in Fig.~\ref{fig:c range}.

\begin{figure*}[!htb]
    \centering
    \subfigure[]
    {
        \includegraphics[width=0.15\textwidth]{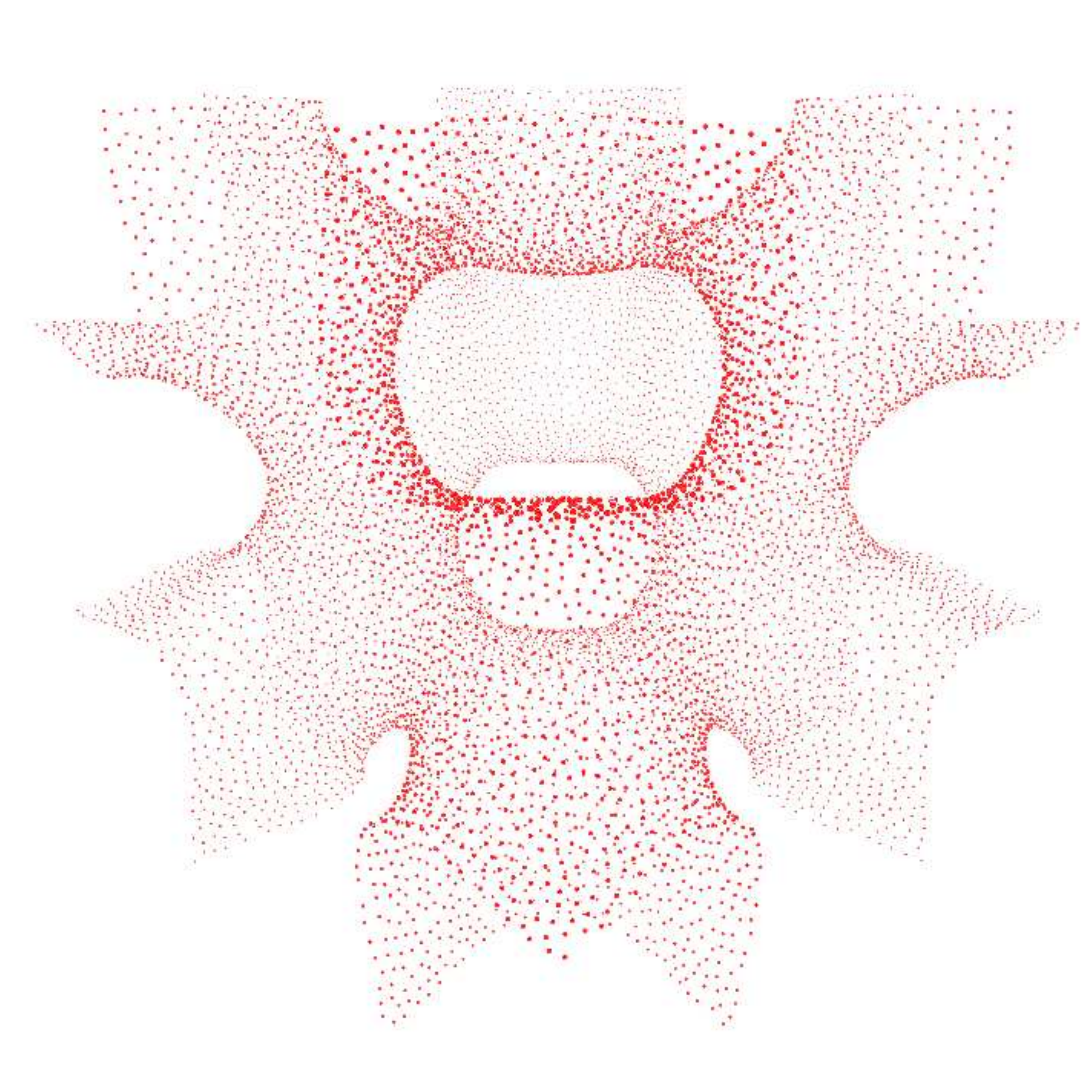}
    }
    \quad	
    \subfigure[]
    { \label{subfig:sheet_IWP_cmp}
        \includegraphics[width=0.26\textwidth]{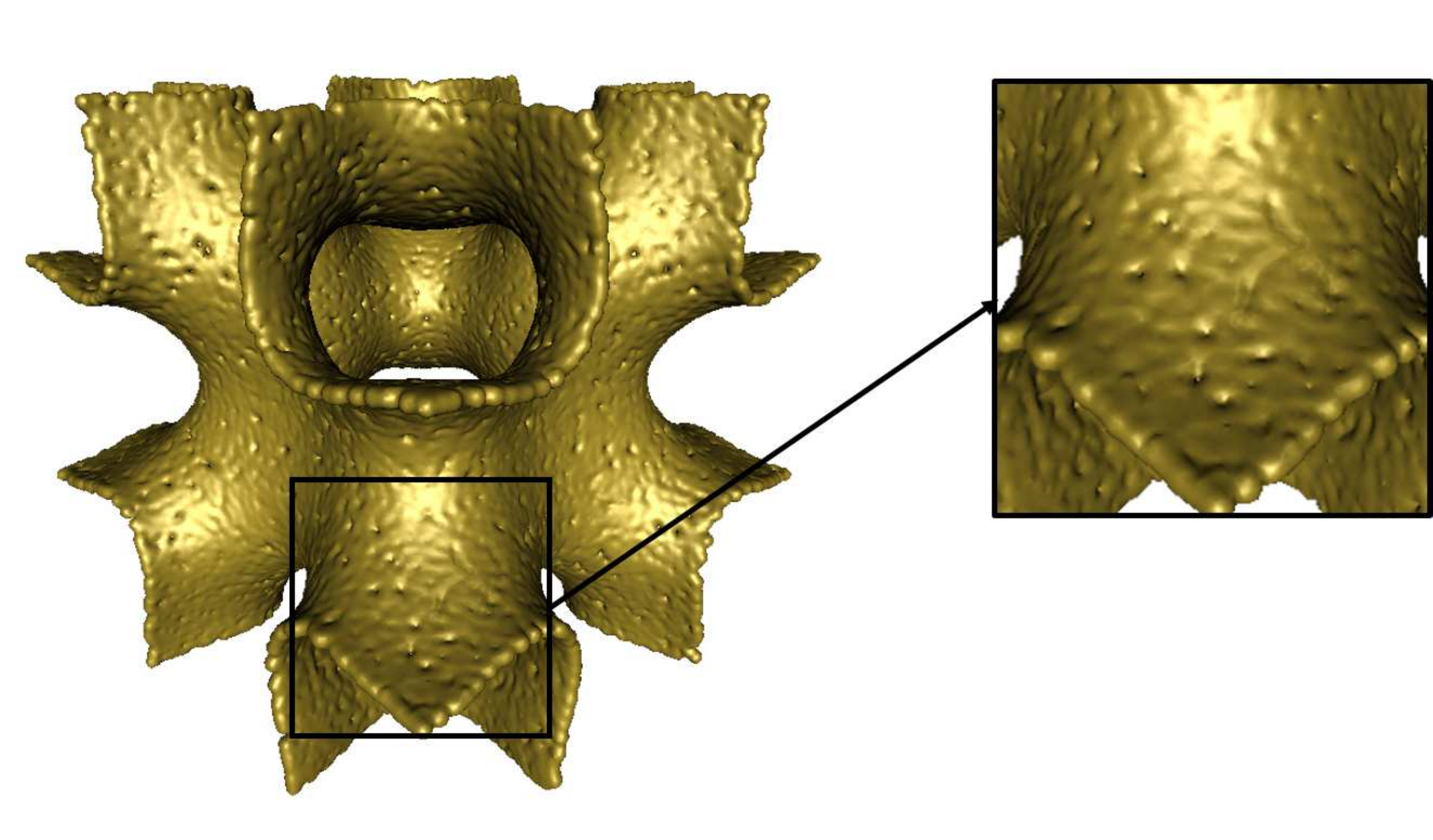}
    }
    \quad
    \subfigure[]
    { \label{subfig:sheet_IWP}
        \includegraphics[width=0.26\textwidth]{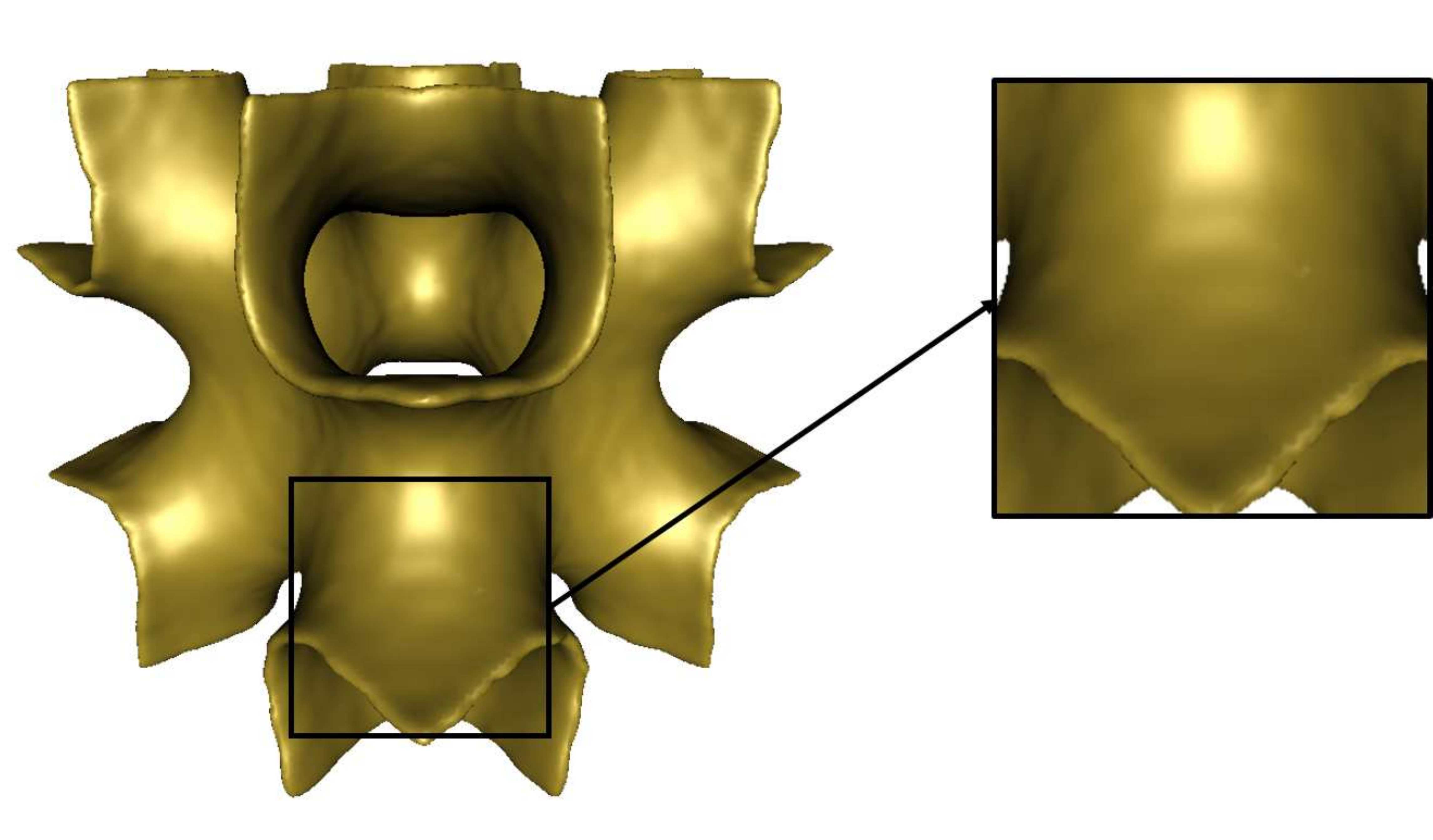}
    }

    \caption
    {
        Comparison with the sheet structure generation method developed in Ref.~\cite{wang2013gpu}.
        (a) Point cloud (\emph{I-WP}) with 10,405 points.
        (b) Porous sheet structure generated by the method
            in Ref. \cite{wang2013gpu} with thickness $0.02$ (the radius of ball is $0.01$),
            where there are lots of local defects.
        (c) Porous sheet structure $f(x,y,z) \leq 0.01$ with
            a $64 \times 64 \times 64$ control grid generated by our method,
            where the thickness is $0.02$,
            and the local defects are eliminated.
    }
    \label{fig:I-WP unit surface}
 \end{figure*}

 Finally, we compare our method with that developed
    in Ref.~\cite{wang2013gpu},
    where sheet structures are generated by computing the union of balls centered at input points.
 When the point cloud density is low and the offset distance is small,
    balls centered at input points with a small offset distance as the radius may not intersect,
    leading to local defects in the generated sheet structures.
 In the porous structure generated by the method in Ref.~\cite{wang2013gpu}
    (Fig.~\ref{subfig:sheet_IWP_cmp}),
    the thickness is $0.02$ (the ball radius is $0.01$),
    and there are a number of defects.
 However, as shown in Fig.~\ref{subfig:sheet_IWP},
    the porous sheet structure with the same thickness $0.02$ generated by our method is much smoother than that by the method in Ref.~\cite{wang2013gpu},
    where the control grid size of the trivariate B-spline function is $64 \times 64 \times64$.

 \begin{figure*}[!htb]
    \centering
    \subfigure[1-PD of the distance field $f(x,y,z)$ and \emph{D} sheet structures $f(x,y,z) \leq c$.]
    {\label{subfig:sheet_D_PD} \includegraphics[width=0.98\textwidth]{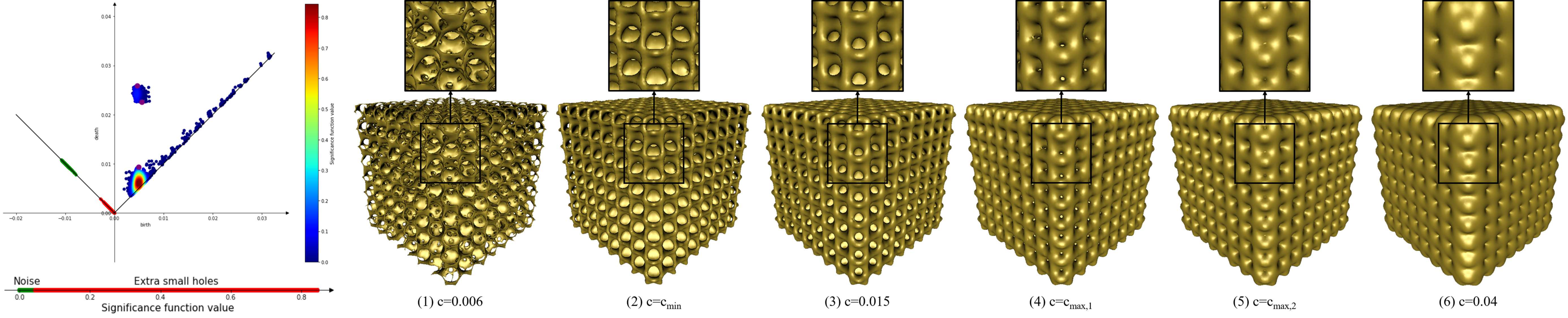}}

    \subfigure[1-PD of the distance field $f(x,y,z)$ and \emph{G} sheet structures $f(x,y,z) \leq c$.]
    {\label{subfig:sheet_G_PD} \includegraphics[width=0.98\textwidth]{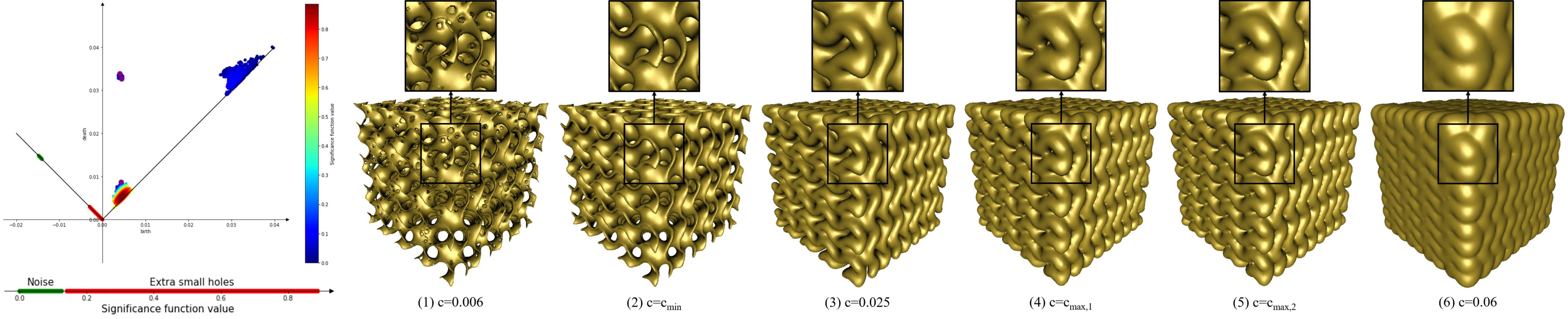}}

    \subfigure[1-PD of the distance field $f(x,y,z)$ and \emph{P} sheet structures $f(x,y,z) \leq c$.]
    {\label{subfig:sheet_P_PD} \includegraphics[width=0.98\textwidth]{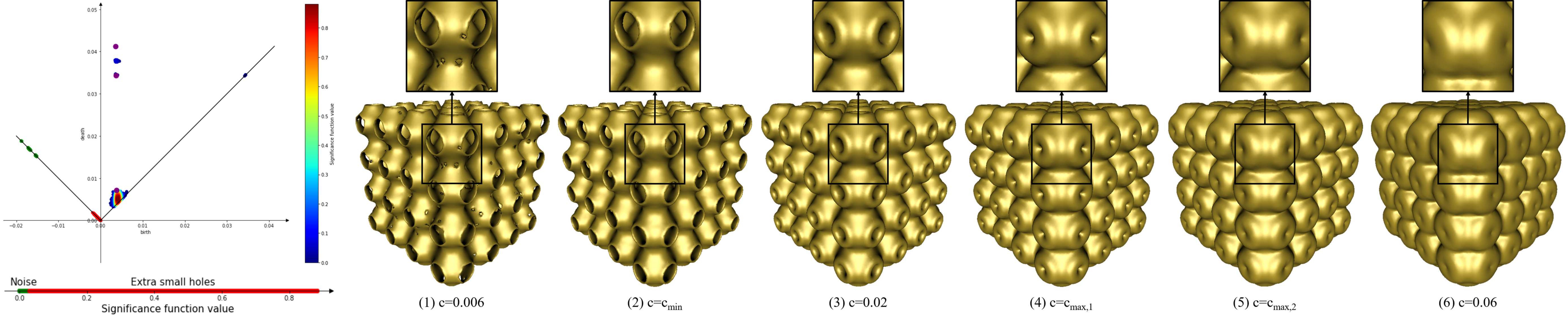}}

    \subfigure[1-PD of the distance field $f(x,y,z)$ and \emph{Venus I-WP} sheet structures $f(x,y,z) \leq c$.]
    {\label{subfig:sheet_venus_IWP_PD} \includegraphics[width=0.98\textwidth]{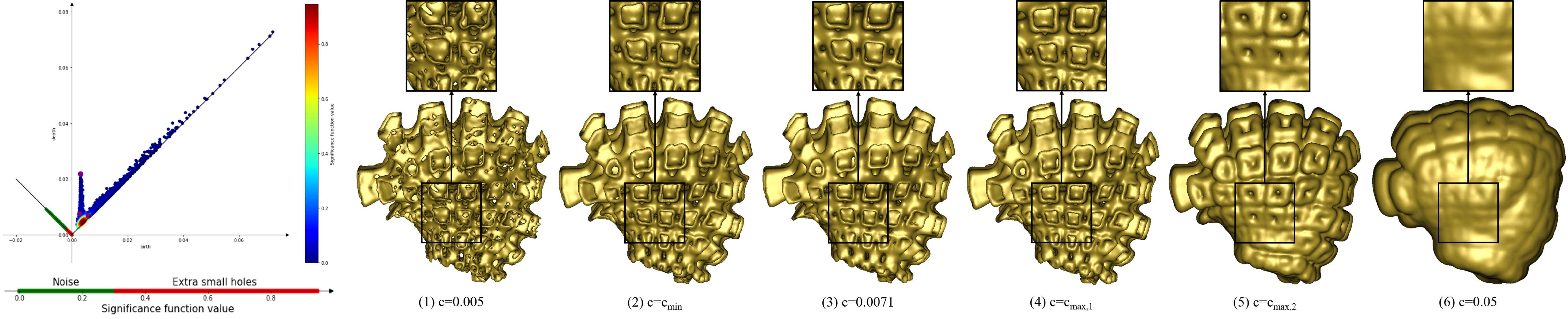}}

    \subfigure[1-PD of the distance field $f(x,y,z)$ and \emph{Balljoint I-WP} sheet structures $f(x,y,z) \leq c$.]
    {\label{subfig:sheet_balljoint_IWP_PD} \includegraphics[width=0.98\textwidth]{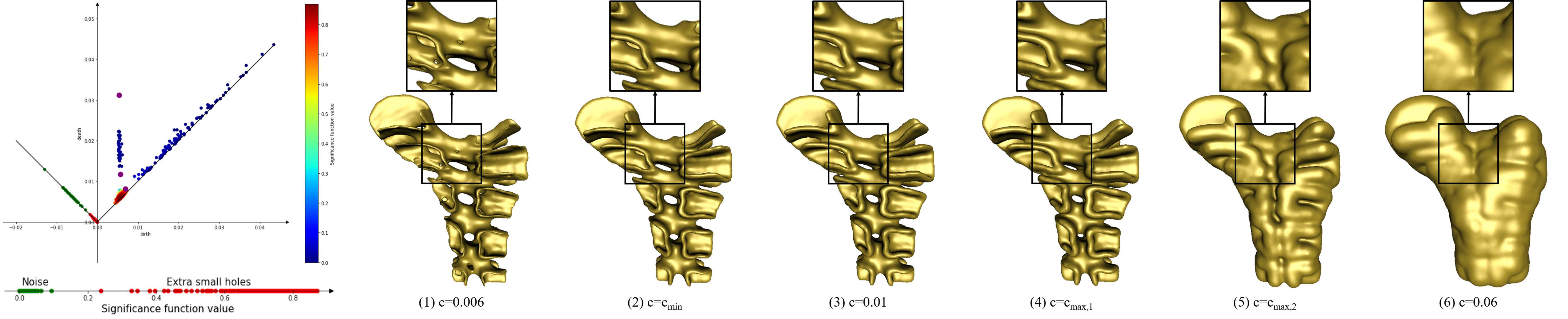}}

    \caption{Generated porous sheet structures $f(x,y,z) \leq c$ with different thickness parameter $c$.}
    \label{fig:Porous c_range}
 \end{figure*}

\subsection{Thickness range determination}
\label{subsec:topology}

 \begin{table*}[!htb]
    \centering
    \caption{The thickness thresholds of different sheet structures.}
    \label{table:thickness}
    \resizebox{0.48\textwidth}{!}{
    \begin{threeparttable}
    \begin{tabular}{ccccc}
    \hline
    Model             & Fig.                                 & ${c}_{min}$               & ${c}_{max,1}$             & ${c}_{max,2}$ \\
    \hline
    D                 & \ref{subfig:sheet_D_PD}              & $9.29 \times {10}^{-3}$   & $2.26 \times {10}^{-2}$   & $2.58 \times {10}^{-2}$ \\
    G                 & \ref{subfig:sheet_G_PD}              & $8.67 \times {10}^{-3}$   & $3.26 \times {10}^{-2}$   & $3.39 \times {10}^{-2}$ \\
    P                 & \ref{subfig:sheet_P_PD}              & $7.20 \times {10}^{-3}$   & $3.43 \times {10}^{-2}$   & $4.13 \times {10}^{-2}$ \\
    Venus I-WP        & \ref{subfig:sheet_venus_IWP_PD}      & $6.77 \times {10}^{-3}$   & $7.49 \times {10}^{-3}$   & $2.18 \times {10}^{-2}$ \\
    Balljoint I-WP    & \ref{subfig:sheet_balljoint_IWP_PD}  & $8.16 \times {10}^{-3}$   & $1.17 \times {10}^{-2}$   & $3.13 \times {10}^{-2}$ \\
    \hline
    \end{tabular}
    \end{threeparttable}
    }
 \end{table*}
 \begin{table*}[!htb]
    \centering
    \caption{The variation of the pore number in the sheet structures $f(x,y,z) \leq c$ in Fig. \ref{fig:Porous c_range}.}
    \label{table:betti}
    \resizebox{0.72\textwidth}{!}{
    \begin{threeparttable}
    \begin{tabular}{ccccccccccc}
    \hline
    Model           & Fig.                                  & \multicolumn{6}{c}{Topological measurement $\bar{\beta}$~\pref{eq:topo_measure}}                                                        \\
    \cline{3-8}
                    &                                       & $c <{c}_{min}$         & $c={c}_{min}$         & ${c}_{min} < c < {c}_{max,1}$
                                                            & $c = {c}_{max,1}$      & $c = {c}_{max,2}$     & $c > {c}_{max,2}$   \\
    \hline
    D               & \ref{subfig:sheet_D_PD}               & 5.661                  & 1.000                 & 1.000
                                                            & 0.998                  & 0.000                 & 0.000   \\
    G               & \ref{subfig:sheet_G_PD}               & 6.171                  & 1.000                 & 1.000
                                                            & 0.998                  & 0.000                 & 0.000   \\
    P               & \ref{subfig:sheet_P_PD}               & 2.681                  & 1.000                 & 1.000
                                                            & 0.953                  & 0.000                 & 0.000   \\
    Venus I-WP      & \ref{subfig:sheet_venus_IWP_PD}       & 4.001                  & 1.000                 & 1.000
                                                            & 0.997                  & 0.000                 & 0.000   \\
    Balljoint I-WP  & \ref{subfig:sheet_balljoint_IWP_PD}   & 1.838                  & 1.000                 & 1.000
                                                            & 0.973                  & 0.000                 & 0.000   \\
    \hline
    \end{tabular}
    \end{threeparttable}
    }
 \end{table*}
 Implicit representation provides flexible freedom to represent porous
    sheet structure with complicated topology structure and geometric shape.
 However, conventional sheet generation methods cannot identify which thickness
    parameter $c$ maintains the pores open~\cite{yoo2009general,hu2020efficient}.
 With the method developed in Section~\ref{subsec:offset distance},
    by persistent homology,
    we can understand and control the topology structure of the porous sheet.
 Specifically, by two round clustering,
    the points in the 1-PD of the distance field $f(x,y,z)$ can be segmented into three clusters, i.e.,
    pore cluster $\mathcal{C}_p$~\eqref{eq:pore_cluster},
    hole cluster $\mathcal{C}_h$~\eqref{eq:hole_cluster},
    and topological noise cluster  $\mathcal{C}_t$~\eqref{eq:noise_cluster}.
 With the pore cluster and hole cluster,
    we can determine three thresholds of the thickness
    parameter $c$, i.e., $c_{max,1}$, $c_{max,2}$~\eqref{eq:c_max}, and $c_{min}$~\eqref{eq:c_min}.
 When $c \in [c_{min},c_{max,1})$,
    the pores of the implicit porous structure $f(x,y,z) \leq c$ remain open;
    when $c \geq c_{max,1}$,
    some pores will be closed;
    when $c \geq c_{max,2}$,
    all pores will be closed.
 As illustrated in Fig.~\ref{fig:Porous c_range},
    five examples are presented to demonstrate the thickness range determination method.
 The three thickness thresholds $c_{max,1}$, $c_{max,2}$~\eqref{eq:c_max}, and $c_{min}$~\eqref{eq:c_min} of different sheet structures are listed in Table~\ref{table:thickness}.

 \begin{figure}[!htb]
    \centering
    \includegraphics[width=0.35\textwidth]{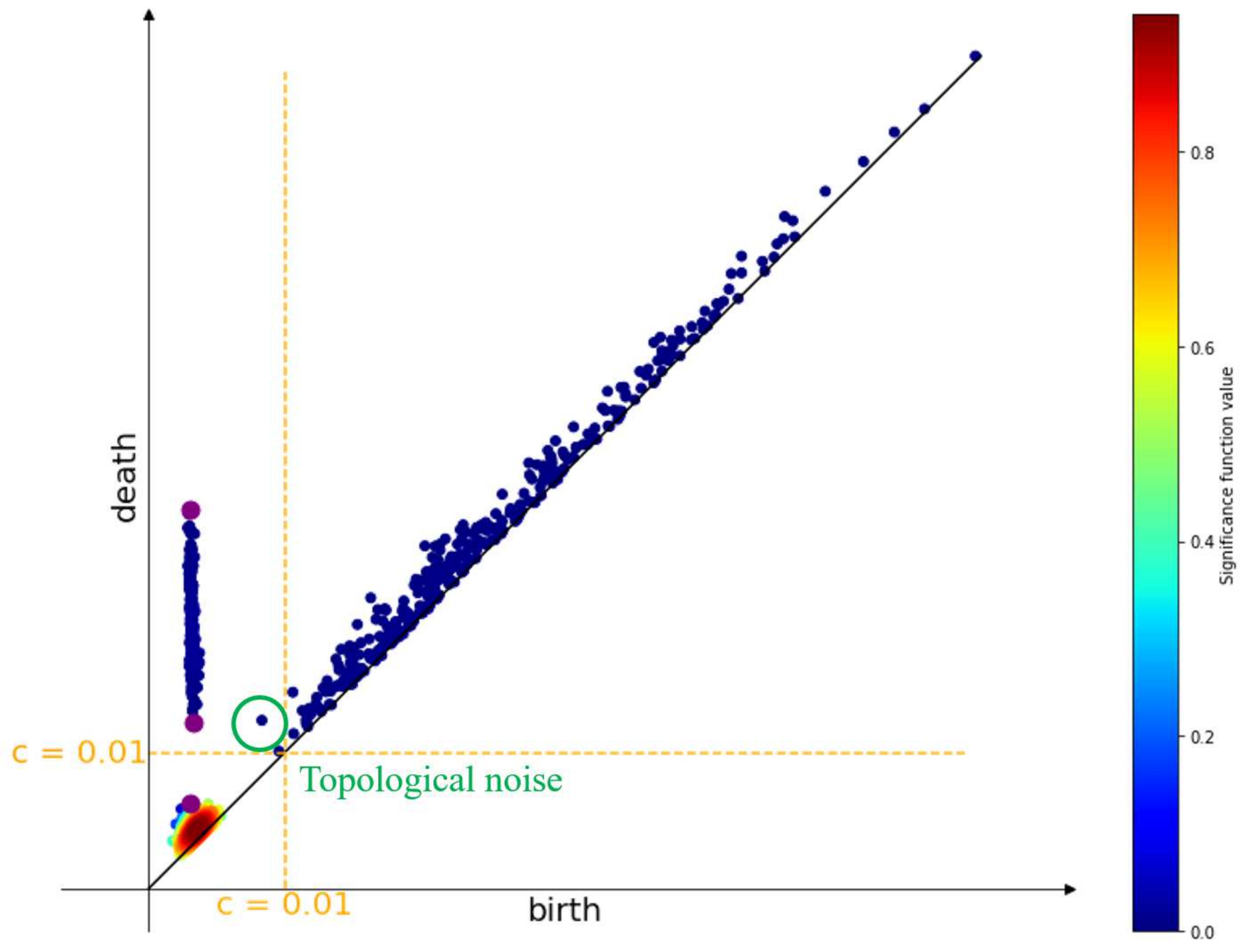}
    \caption
    {
        1-PD of distance field $f(x,y,z)$ (\emph{Tooth P}),
            where the number of points $\{(b_i, d_i) \in  \text{1-PD} | b_i \leq 0.01 < d_i\}$
            is the 1-Betti number of the sheet structure $f(x,y,z) \leq 0.01$.
    }
    \label{fig:tooth_P_betti}
 \end{figure}
 As shown in Fig.~\ref{fig:tooth_P_betti},
    the 1-Betti number of the sheet structure $f(x,y,z) \leq c$ is the number of points
    $\{(b_i, d_i) \in  \text{1-PD} | b_i \leq c < d_i\}$,
    which possibly contain points in the pore cluster, hole cluster,
    and topological noise cluster.
 Therefore, the number of pores and holes
    in the sheet structure $f(x,y,z) \leq c$ can be represented as
    \begin{equation}
       B(c)=\beta - {{\beta }_{t}},
    \end{equation}
    where ${\beta }$ is the 1-Betti number of the sheet structure,
    and ${{\beta }_{t}}$ is the number of points
    $\{(b_i^t, d_i^t) \in  \mathcal{C}_t | b_i^t \leq c < d_i^t\}$
    in the topological noise cluster.
 To measure the variation of the number of pores and holes in the
    sheet structure $f(x,y,z) \leq c$,
    we define the following \emph{topological measurement}
    \begin{equation} \label{eq:topo_measure}
        \bar{\beta }=\frac{B(c)}{B(c_{min})}.
    \end{equation}

 The topological measurements of the five examples are listed in
    Table \ref{table:betti}.
 When $c = c_{min}$, the topological measurements $\bar{\beta}$ are all equal to $1$.
 When $c < c_{min}$, the topological measurements $\bar{\beta}$ are greater than $1$
    because of the extra small holes in the pore sheet structure $f(x,y,z) \leq c\ (c < c_{min})$.
 When the thickness parameter $c \in [{c}_{min}, {c}_{max,1})$,
    the topological measurements of the sheet structures $f(x,y,z) \leq c$ remain unchanged as $1$,
    indicating that the pores in the sheet structures remain open.
 Moreover, when the thickness parameter $c \geq {c}_{max,1}$,
        the topological measurements start to decrease,
        which means some pores are closed.
 Finally, when the thickness parameter $c \geq {c}_{max,2}$,
        the topological measurements are all equal to $0$,
        indicating that all pores are closed.
 These examples verify the theoretical analysis
    in Section~\ref{subsec:offset distance},
    and validate the effectiveness of the developed method.

\subsection{Efficiency of direct slicing}
\label{subsec:Eff direct}

 \begin{table*}[!htb]
    \centering
    \caption{The statistics on slicing sheet structures using traditional method and our method.}
    \label{table:slice}
    \resizebox{0.9\textwidth}{!}{
    \begin{threeparttable}
    \begin{tabular}{cccccccc}
    \hline
    Model       & Thickness parameter $c$   &Volume ratio   & STL Size (MB)\tnote{1}   & Resolution\tnote{2}       & Layers   & \multicolumn{2}{c}{Slicing time   (seconds)} \\ \cline{7-8}
                &                           &               &                 &                           &          & Traditional method \tnote{3}                   & Our method     \\
    \hline
    D           & 0.021                     & 75\%          & 97.6            & $100\times100\times100$   & 74       & 304.18    & 3.74     \\
                &                           &               & 390             & $200\times200\times200$   & 148      & 1,351.74  & 29.08    \\
    G           & 0.02                      & 60\%          & 92.2            & $100\times100\times100$   & 73       & 125.84    & 3.70     \\
                &                           &               & 373             & $200\times200\times200$   & 148      & 561.48    & 30.01    \\
    P           & 0.022                     & 50\%          & 76.3            & $100\times100\times100$   & 74       & 138.07    & 3.52     \\
                &                           &               & 302             & $200\times200\times200$   & 148      & 583.01    & 26.81    \\
    Tooth P     & 0.01                      & 20\%          & 42.1            & $100\times100\times100$   & 64       & 42.23     & 3.24     \\
                &                           &               & 169             & $200\times200\times200$   & 128      & 199.19    & 22.67    \\
    Venus I-WP  & 0.008                     & 20\%          & 40.0            & $100\times100\times100$   & 68       & 76.18     & 3.14     \\
                &                           &               & 161             & $200\times200\times200$   & 137      & 328.24    & 23.67    \\
    \hline
    \end{tabular}
    \begin{tablenotes}
    \footnotesize
    \item[1] The storage space of triangle mesh using the traditional STL file format.
    \item[2] The resolution of the distance field for the MC algorithm and the MS algorithm.
    \item[3] The slicing time of the traditional method including the time to generate and slice the triangular mesh.
    \end{tablenotes}
    \end{threeparttable}
    }
 \end{table*}

 The implicit B-spline represented porous sheet structure can be sliced directly and
    efficiently.
 Slicing is the process of transforming an input model into a series of contours
    that provide the printing area for filling material.
 The slice file is a layer-by-layer representation of a model,
    where each layer consists of some disjoint polygons.
 For the implicit sheet structures $f(x,y,z) \leq c$,
    the contours on the slicing plane $z = z_{0}$ can be expressed as $f(x,y,z_{0})=c$.
 Using the MS algorithm, 
    we can directly slice the implicit porous sheet structures and extract closed contours.

 We compare the direct slicing method for implicit porous sheet structure
    with the traditional slicing method,
    similar to the comparison strategy adopted in Ref.~\cite{song2018function}.
 For this purpose, we initially transform the implicit porous sheet structure
    into a triangular mesh model,
    using the marching cubes (MC) algorithm~\cite{lorensen1987marching}.
 Then, by the contour extraction method developed
    in Ref.~\cite{lin2013cutting},
    we can slice the triangular mesh model and acquire the contours.

 The statistics of slicing based on triangular mesh (\emph{traditional method})
    and implicit B-spline representation (\emph{our method}) are listed in Table~\ref{table:slice},
    where the implicit porous sheet structures are generated with the method presented in Section~\ref{subsec:selection}.
 In Table~\ref{table:slice},
    the time of the traditional method includes the time to generate and slice the triangular mesh model.
 The time of the traditional method is at least 8.79 times more than that of our method.
 Moreover,
    the traditional method suffers from the complexity of the internal structure of the model.
 Based on the distance fields of the same resolution of $200 \times 200 \times 200$,
    the time of slicing \emph{D} sheet structure using the traditional method is 2.41 times more than that of \emph{G} sheet structure (refer to Table~\ref{table:slice}),
    whereas the time of our method is nearly the same.
 In addition,
    as the resolution of the distance field increases,
    the triangular meshes generated from porous sheet structures become increasingly difficult to store.
 In comparison,
    by storing the control grid size and control coefficients,
    the storage space of five sliced sheet structures
    with the $128\times128\times128$ control grid
    in Table~\ref{table:slice} is $20.0$ MB.

\begin{figure*}[!htb]
    \centering
    \includegraphics[width=0.85\textwidth]{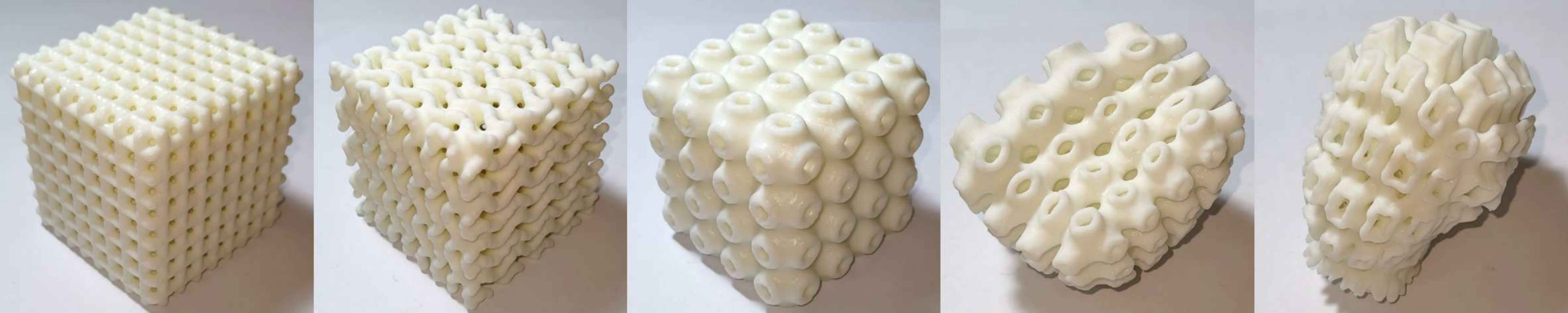}
    \caption{3D printed porous sheet structures with the SLA technology. }
    \label{fig:AM sheet}
 \end{figure*}

 The contours acquired by the direct slicing can be saved as commonly used layer files,
    such as SLC files and projection images,
    which can be directly used in 3D printing because implicit sheet structures are naturally closed.
 In this experiment, all sliced sheet structures are manufactured with SLA technology.
 As shown in Fig.~\ref{fig:AM sheet},
    all sliced porous sheet structures are correctly manufactured,
    thereby illustrating that the proposed sheet generation method can be effectively used in 3D printing.

\subsection{Limitations and future work}
\label{subsec:limitation}

 The developed method in this paper can generate the implicit
    sheet structure from the point cloud
    and control its topology structure.
 However, it has several limitations.
 When the size of the pores on porous surfaces is relatively small,
    topological noise with long persistence times
    will probably affect the clustering of points in 1-PD.
 When the persistence times of topological noise are longer than those of some pores,
    the determination of the thickness thresholds ${c}_{max,1}$, ${c}_{max,2}$~\eqref{eq:c_max}
    and ${c}_{min}$~\eqref{eq:c_min} will be affected.

 In the future,
    we will consider a better pre-processing method for extracting topological noise that considers geometric and topological information.

\section{Conclusion}
\label{sec:conclusion}

 In this work,
    we propose a method for generating porous sheet structures
    from point clouds.
 The generated sheet structure has
    an implicit B-spline representation,
    which facilitates the representation and storage of geometrically
    and topologically complex models.
 Compared with traditional sheet structure generation methods,
    the generation of porous sheet structures in this paper
    does not require redundant closure operations,
    and the generated sheet structures can ensure smoothness.
 It has high-quality generation results for low-density and noisy point clouds.
 Moreover, compared with previous research,
    we analyze the reasonable thickness range of the generated sheet structures for the first time.
 Within the derived thickness range,
    we can control the topology of the generated sheet structure according to the requirements.
 The generated sheet structures can be sliced directly with the MS algorithm,
    and generated contours can be used directly in 3D printing.
 Compared with the conventional slicing based on the triangular mesh,
    experimental results demonstrate that slicing based on the implicit B-spline representation generated in this research reduces computation and storage costs effectively,
    which is conducive to industrial manufacturing.

\section*{References}
\bibliography{mybibfile}

\end{document}